# Endo-exo framework for a unifying classification of episodic landslide movements

Qinghua Lei[1,*], Didier Sornette[2]

[1]*Department of Earth Sciences, Uppsala University, Uppsala, Sweden*

[2]*Institute of Risk Analysis, Prediction and Management, Academy for Advanced Interdisciplinary Studies, Southern University of Science and Technology, Shenzhen, China*

[*]Corresponding author: qinghua.lei@geo.uu.se

**Abstract**

Landslides exhibit intermittent gravity-driven downslope movements developing over days to years before a possible major collapse, commonly boosted by external events like precipitations and earthquakes. The reasons behind these episodic movements and how they relate to the final instability remain poorly understood. Here, we develop a novel "endo-exo" theory to quantitatively diagnose landslide dynamics, capturing the interplay between exogenous stressors such as rainfall and endogenous damage/healing processes. We predict four distinct types of episodic landslide dynamics (endogenous/exogenous-subcritical/critical), characterized by power law relaxations with different exponents, all related to a single parameter $\vartheta$. These predictions are tested on the dataset of the Preonzo landslide, which exhibited multi-year episodic movements prior to a catastrophic collapse. All its sporadic activities can be accounted for within this classification with $\vartheta \approx 0.45 \pm 0.1$, providing strong support for our parsimonious theory. We find that the final collapse of this landslide is clearly preceded over 1-2 months by an increased frequency of medium/large velocities, signaling the transition into a catastrophic regime with amplifying positive feedbacks. Our research suggests that landslides may not permanently operate at a critical state, which has major implications for forecasting catastrophic failure events.

**Main text**

Landslides, a widespread form of mass wasting, occur in various Earth surface environments and pose significant threats to life and property worldwide[1–4]. Due to rapid population growth and urbanization, human habitats are increasingly exposed to landslide hazards, with the situation becoming even more severe under climate change, where extreme rainfall, permafrost thaw, and glacier retreat have promoted fatal landslides[5]. Extensive field observations show that landslides commonly exhibit episodic movements characterized by intermittent acceleration-deceleration sequences that are boosted by external events like precipitations and earthquakes[6–14]. Some landslides have episodically moved over hundreds or thousands of years without collapse, while others could evolve into a major collapse after episodically deforming over days to years[4]. The reasons behind these episodic movements (marked by intermittent bursts of displacement activities followed by sustained periods of relaxation dynamics) and how they relate to a possible final catastrophic failure remain poorly understood, inhibiting our capability to predict landslide behavior and mitigate the associated risks.

We identify the following fundamental questions: (a) Are episodic landslide movements of an exogenous or endogenous origin? (b) What are their underlying mechanisms? (c) How do they relate to catastrophic failures? Here, by exogeneous (exo), we refer to external forcing like rainfall, snowmelt, seismicity, and anthropogenic activities; by endogenous (endo), we refer to internal processes such as damage, healing, internal faulting, and evolution of frictional sliding properties. Fig. 1 presents two typical examples of episodic landslide movements, illustrating the differences in precursory/recovery dynamics around intermittent velocity peaks of exogenous versus endogenous origin. The Vallcebre landslide (Spain)[23] displayed an exogenous velocity peak (associated with rainfall-induced groundwater table fluctuation), characterized by a sudden jump followed by a gradual recovery that exhibits a power law trend (Fig. 1a). In contrast, the La Clapière landslide (France)[24] experienced progressive velocity growth, culminating in an endogenous peak, and then an almost symmetrical decay, with both the precursory and recovery dynamics approximately following the same power law trend (Fig. 1b). Similar episodic movement patterns are observed in various landslides across different types, including rockfalls, rockslides, soilslides, and earthflows (see Supplementary Note 1, Supplementary Figs. 1 and 2, and Supplementary Table 1). We establish a novel "endo-exo" theoretical framework to quantitatively diagnose the precursory and/or recovery properties of such intermittent velocity peaks. The rationale behind is that a complex system (like a landslide) consisting of many interacting components can exhibit specific precursory/recovery patterns in response to significant fluctuations in its activity[15], providing signatures for distinguishing an endogenous origin from an exogenous cause, as illustrated here for landslides (Fig. 1 and Supplementary Figs. 1 and 2) and previously demonstrated in diverse contexts including earthquake sequences[16,17], book sales[18], YouTube views[19], financial markets[20], social shocks[21], and human diseases[22].

Actually, many natural and social complex systems are subject to the similar dual effect of external perturbation and internal organization, for which the endo-exo question is relevant, including earthquake faults, volcanoes, engineering structures, financial markets, electrical grids, social communities, traffic networks, and biological organisms, among others[25]. In a nutshell, all these systems, which are out-of-equilibrium (i.e., with macroscopic properties changing over time) and open (i.e., coupled to the outside), self-organize as a result of the interplay between exogenous stressors (e.g., shocks, influx of energy/matter) and endogenous processes (e.g., degradation, restoration, adaptation). Our foundational endo-exo theory provides a general and quantitative framework to understand the evolutionary behavior of complex systems[25]. This framework differentiates between endogenous factors—internal mechanisms driving system behavior—and exogenous influences—external shocks or interventions impacting the system as well as entangled situations where different constituents of the system strongly interact resulting in mingled causality. We emphasize that the broad applicability of the endo-exo theory across diverse contexts is not based on a superficial analogy but rather on the fundamental universality of the endo-exo concept and the generality of its mathematical formulation.

In the current paper, in the context of landslides, we focus on the regime of episodic movements, where their endo/exo origins (i.e., initial triggering sources) can be distinguished. We provide a thorough demonstration of our theory based on the long-term monitoring dataset of a rainfall-induced landslide at Preonzo, Switzerland, which episodically moved over many years prior to a major collapse. We also explore the mechanisms underlying the transition of this landslide from episodic to catastrophic regimes, prior to the final collapse. Our findings open up a new avenue for understanding and predicting catastrophic slope failures.

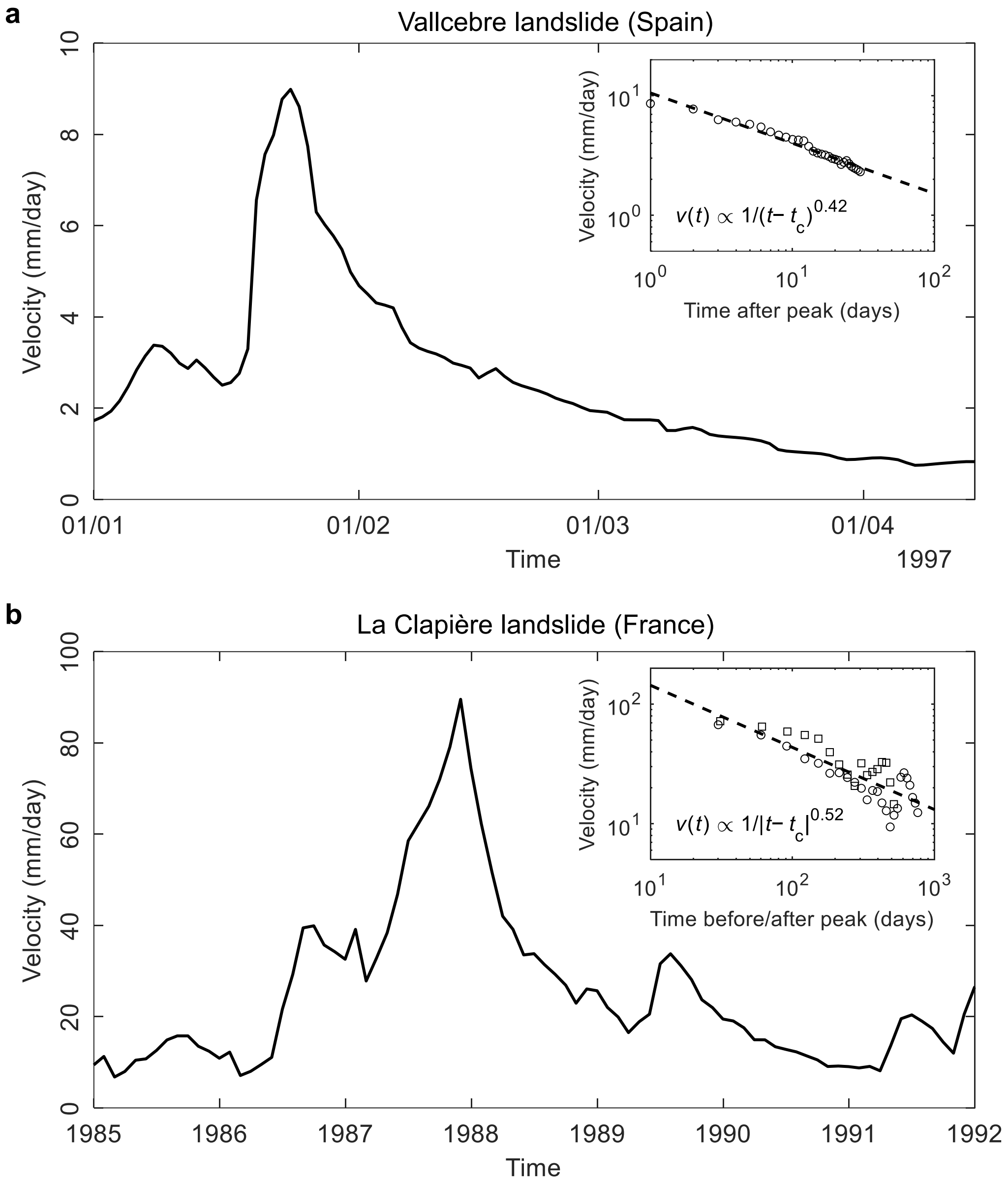

**Fig. 1 Typical episodic landslide movement behavior around a major velocity peak. a** Velocity time series of the Vallcebre landslide (Spain) around an exogeneous velocity peak (associated with rainfall-induced groundwater table fluctuation); inset shows the power law velocity dynamics $v(t)$ after the peak at time $t_c$. **b** Velocity time series of the La Clapière (France) around an endogenous velocity peak; inset shows the power law velocity dynamics $v(t)$ before (squares) and after (circles) the peak at time $t_c$.

## Results

**An epidemic cascade model of mass interactions.** We conceptualize a landslide as a complex system consisting of numerous geomaterial masses lying on an inclined substrate and interacting via cohesive or frictional contacts (Fig. 2). Such a conceptual picture resembles the spring-slider model, which was originally developed for earthquakes[26] and later demonstrated to be also applicable for landslides[27]. Geomaterial masses here could be envisioned as different components or blocks within an active landslide, with the number of masses assumed to be large. We establish a theoretical framework based on the principles of statistical physics and derive analytical solutions to describe the macroscopic behavior of such a complex mass system.

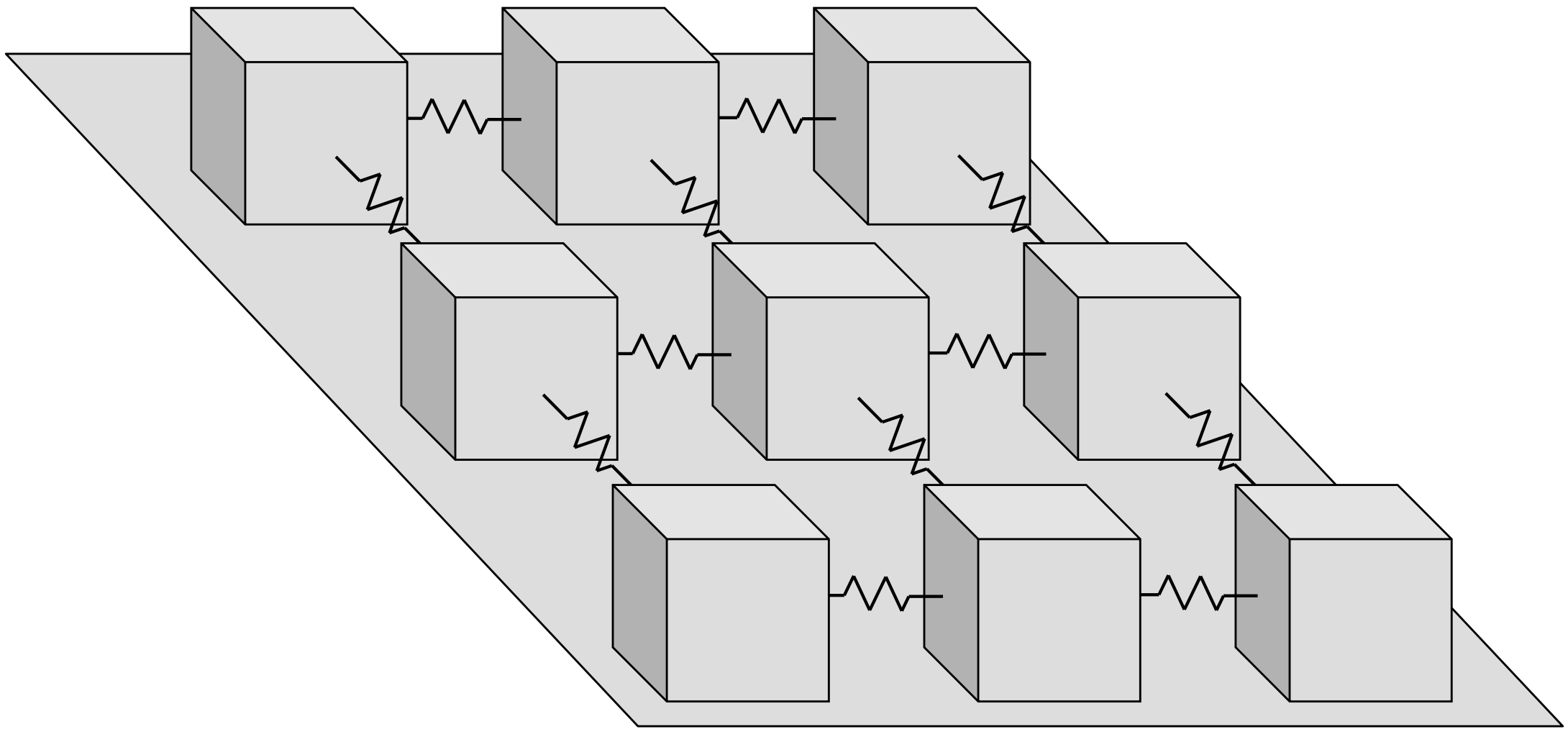

**Fig. 2 Conceptual picture of a landslide.** The landslide as a complex system consists of numerous geomaterial masses (represented by sliders) lying on an inclined substrate and interacting via cohesive or frictional contacts (represented by springs).

The displacement activity of a landslide results from a combination of external forces like precipitations and earthquakes in the presence of gravity, and of internal influences where each past moved mass may prompt other masses in its network of interactions to move as a result of the redistribution of mechanical stress, pore pressure, and possibly other physico-chemical fields. Starting from an initial moved mass, i.e., the "mother" mass, which first moves due to either external forces or internal fluctuations, it may trigger the movements of first-generation "daughter" masses nearby, which themselves trigger their own daughter masses to move, and so on. Such an epidemic process of mass interactions in the system with a large number of masses is governed by[15,25]:

$$v(t) = V(t) + n \int_{-\infty}^{t} \psi(t-\tau)\, v(\tau) d\tau = V(t) + n \int_{-\infty}^{t} V(\tau) \Psi(t-\tau)\, d\tau, \quad (1)$$

where $v(t)$ is the average displacement rate (i.e., velocity) and $V(t)$ represents the exogenous activation that is not triggered by any epidemic effect in the system (it can be understood as capturing the effects of strong external events that trigger sudden velocity responses in the system). Parameter $n \geq 0$ is the so-called branching ratio, defined as the average number of triggered first-generation daughter masses per mother mass. For $n \leq 1$, it is also the fraction of triggered mass movements over all mass movements; for instance, $n \approx 0.61$ (a scenario analyzed later) implies that ~61% of mass movements are endogenously triggered and ~39% are exogenously caused. $\psi(t-\tau)$ is the probability distribution function of the waiting time between the movement of a mother mass at time $\tau$ and the movement of a first-generation daughter mass at a later time $t$, while $\Psi(t-\tau)$ is the probability of the waiting time between the movement of a mass at time $\tau$ and the triggered movement of another mass at a later time $t$ through any possible generation lineage. In other words, $\psi(t-\tau)$ governs the triggering within each individual mother-daughter pair, while $\Psi(t-\tau)$ considers the cascades through the full genealogy of mother-daughter descendants. The first equality in equation (1) expresses the fact that the present velocity $v(t)$ is influenced by all previous motions mediated to the present via $\psi(t-\tau)$ that governs each individual interaction, with the resulting equation having a self-consistent integral structure. The second equality in equation (1) provides the formal solution of the integral equation, showing that $v(t)$ results from the past exogenous activation $V(\tau)$ mediated to the present by $\Psi(t-\tau)$ cumulatively incorporating all the generations of cascades of mass interactions.

We assume $\psi(t-\tau)$ obeys a power law characterizing a long-memory process[15]:

$$\psi(t-\tau) \propto 1/(t-\tau)^{1+\vartheta}, \text{ with } 0 < \vartheta < 1 \text{ and for } t-\tau > c \tag{2}$$

where $\vartheta$ controls the persistence of memory (the larger $\vartheta$ is, the shorter the memory is); $c$ is a small characteristic time scale defining the onset of the power law decay and reflecting the rupture time[28] that may arise from brittle creep[29], viscous deformation[30], pore pressure diffusion[31], and/or frictional slip[32]. Here, the condition $\vartheta > 0$ ensures that $\psi(t-\tau)$ is normalizable, while the condition $\vartheta < 1$ is prescribed based on the documented empirical data in the literature[16,33–37] showing that $\vartheta$ rarely exceeds unity. The power law form of $\psi(t-\tau)$, supported by many empirical observations such as Andrade's law of material creep[38] and Omori's law of aftershock activity[34], can be theoretically derived from the constitutive laws of subcritical crack growth[39], rate-state friction[32], and material rheology[33].

The branching ratio $n$ quantifies the degree of endogeneity of how much past mass movements trigger future mass movements. If $n < 1$, the system is in the subcritical regime[40,41], where there is less than one mass triggered per moved mass on average, such that the number of trigged masses eventually decays to zero; in this regime, the system is dissipative, with the energy released by moving masses smaller than the energy lost for triggering them. If $n > 1$, the system is in the supercritical regime[40,41], where the number of trigged masses on average grows exponentially with time[16] or even faster[42]; in this regime, the released energy from a mobilized mass is in general greater than the dissipated energy for mobilizing it. The critical regime of $n \simeq 1$, lying at the borderline between the subcritical and supercritical regimes, has an overall conservation of the number of triggered masses; correspondingly, the system is in a balanced condition of energy release and dissipation. In this paper, we mainly focus on the subcritical and critical regimes with $n \lesssim 1$ to ensure stationarity, whereas the transition into the supercritical regime $n > 1$ related to the emergence of a catastrophic failure[16,42] will be explored in the Discussion section. Note that the above subcritical/critical/supercritical regimes of spatiotemporal cascading behavior should not be confused with the subcritical/critical regimes defined in the context of fracture mechanics-based slope stability analysis[43], where a landslide is treated as a single global mass and the subcritical/critical state depends on the relative magnitudes of the acting stress and available strength. In contrast, our classification of landslide episodic movements is based on the many-body effects of numerous local masses interacting across spatiotemporal scales within the landslide.

**Classification of episodic landslide dynamics.** According to the analytical solutions (i.e., equations (8), (12), and (13) in Methods) of equation (1), landslide velocities around a peak at time $t_c$ can be described by a generalized finite-time singularity power law as:

$$v(t) \propto 1/|t-t_c|^p, \tag{3}$$

where $t_c$ is the critical time chosen as the time of velocity peak, while the exponent $p$ depends on the parameter $\vartheta$ and the regime delineated by a characteristic time $t^*$ (defined by equation (9) in Methods). This allows us to classify episodic landslide movements into four fundamental types (Fig. 3) based on the origin of disturbance (endogenous versus exogenous) and the level of criticality (subcritical or critical):

- Type I: Exogenous-subcritical, with $p = 1 + \vartheta$ for $t - t_c > t^*$ (see equation (8) in Methods). Here, the cascading propensity is limited ($n < 1$), meaning that the exogenously induced velocity burst at time $t_c$ does not cascade beyond the first few generations of triggered masses (Fig. 3a).

- Type II: Exogenous-critical, with $p = 1 - \vartheta$ for $c < t - t_c < t^*$ (see equation (8) in Methods). Here, the system is in a critical state ($n \simeq 1$), such that the exogenously induced velocity peak at time $t_c$ cascades through the system of interconnected masses, triggering the motion of neighboring masses that further trigger their own neighboring masses and so on (Fig. 3b).
- Type III: Endogenous-subcritical, with $p = 0$ for $|t - t_c| > t^*$ (see equation (13) in Methods). The displacement activity does not result from an exogenous event but instead from endogenous interactions in the presence of small external noisy perturbations. Limited cascading occurs ( $n < 1$ ) such that the (small) peak has no apparent precursory/recovery signatures (Fig. 3c).
- Type IV: Endogenous-critical, with $p = 1 - 2\vartheta$ for $c < |t - t_c| < t^*$ (see equation (12) in Methods). The displacement activity originates from endogenous growth/interaction within the system in a critical state ($n \simeq 1$), where the triggering cascades produce an approximately symmetrical power law acceleration-deceleration behavior around the peak (Fig. 3d).

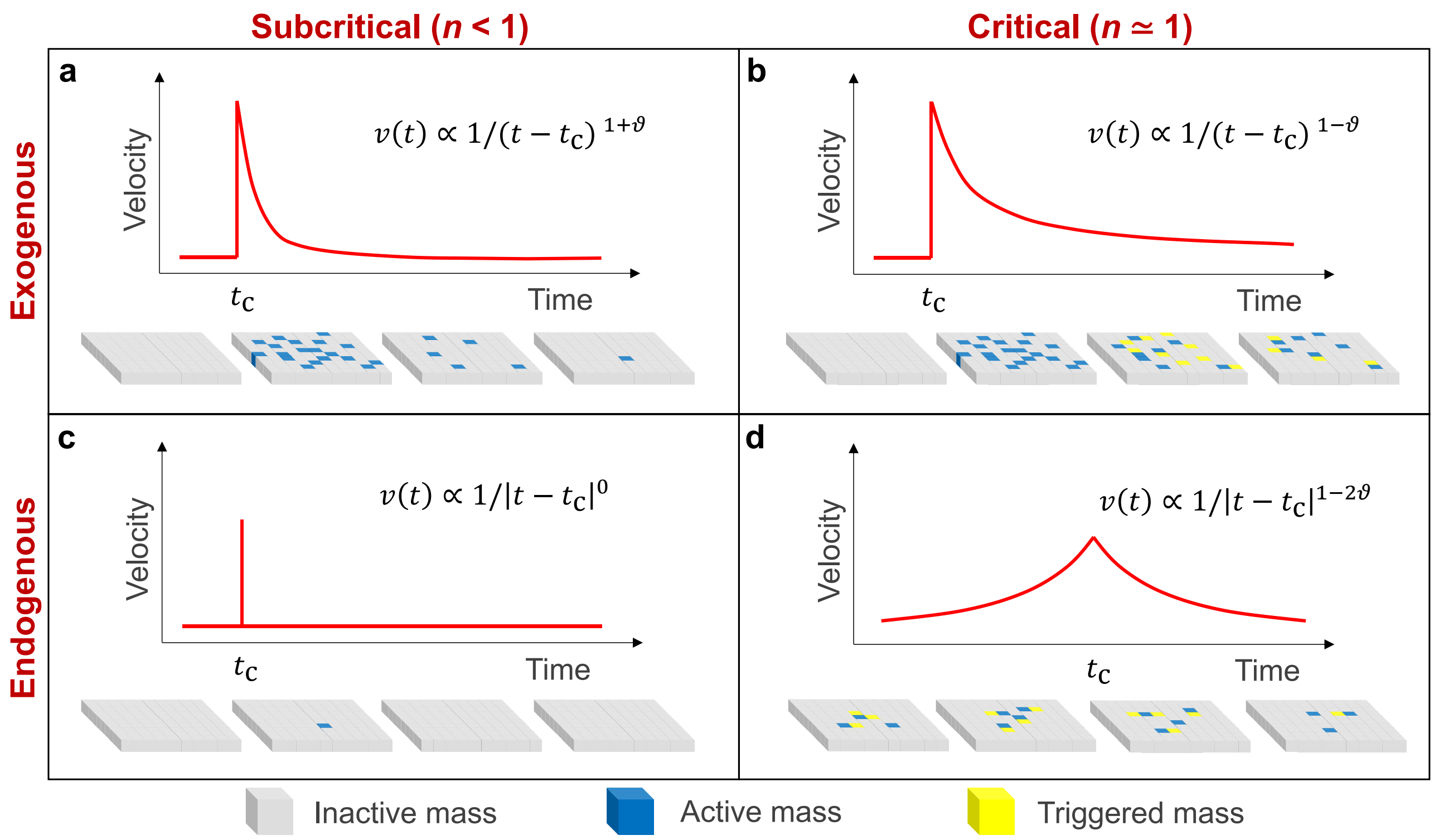

**Fig. 3 Classification of episodic landslide movements based on the origin of disturbance (endogenous versus exogenous) and the level of criticality (subcritical or critical).** There are four distinct types of power law velocity dynamics $v(t)$ around a peak at time $t_c$ and they are all related to a single parameter $\vartheta$. The series of snapshots below each velocity trajectory illustrates the evolution of a mass system (with inactive masses, active masses, and triggered masses color-coded) underpinning the velocity time history.

This classification arises from the interplay of the long-memory process (governing two-mass interactions) as embodied in equation (2) and the epidemic cascade throughout the system (involving many-mass interactions) as captured by equation (1). It can be seen that the relaxation following an endogenous-critical peak (with a smaller exponent $p = 1 - 2\vartheta$) is slower than that following an exogenous-critical peak (with a larger exponent $p = 1 - \vartheta$). This longer-lived influence of an endogenous-critical peak results from the precursory process that impregnates the system much more than its exogenous counterpart[25]. Note that exogenous peaks lack apparent precursors, because they are caused by external events that act as sudden shocks to the system. The distinct properties of different peak types enable differentiation between endogenous and exogenous origins of episodic landslide movements, based on the system's response around peaks rather than (often elusive) correlations with external events. It is important to clarify that the differentiation here is about the initial triggering source (whether internal or external), rather than the drive of the system's response, which is governed by endogenous processes.

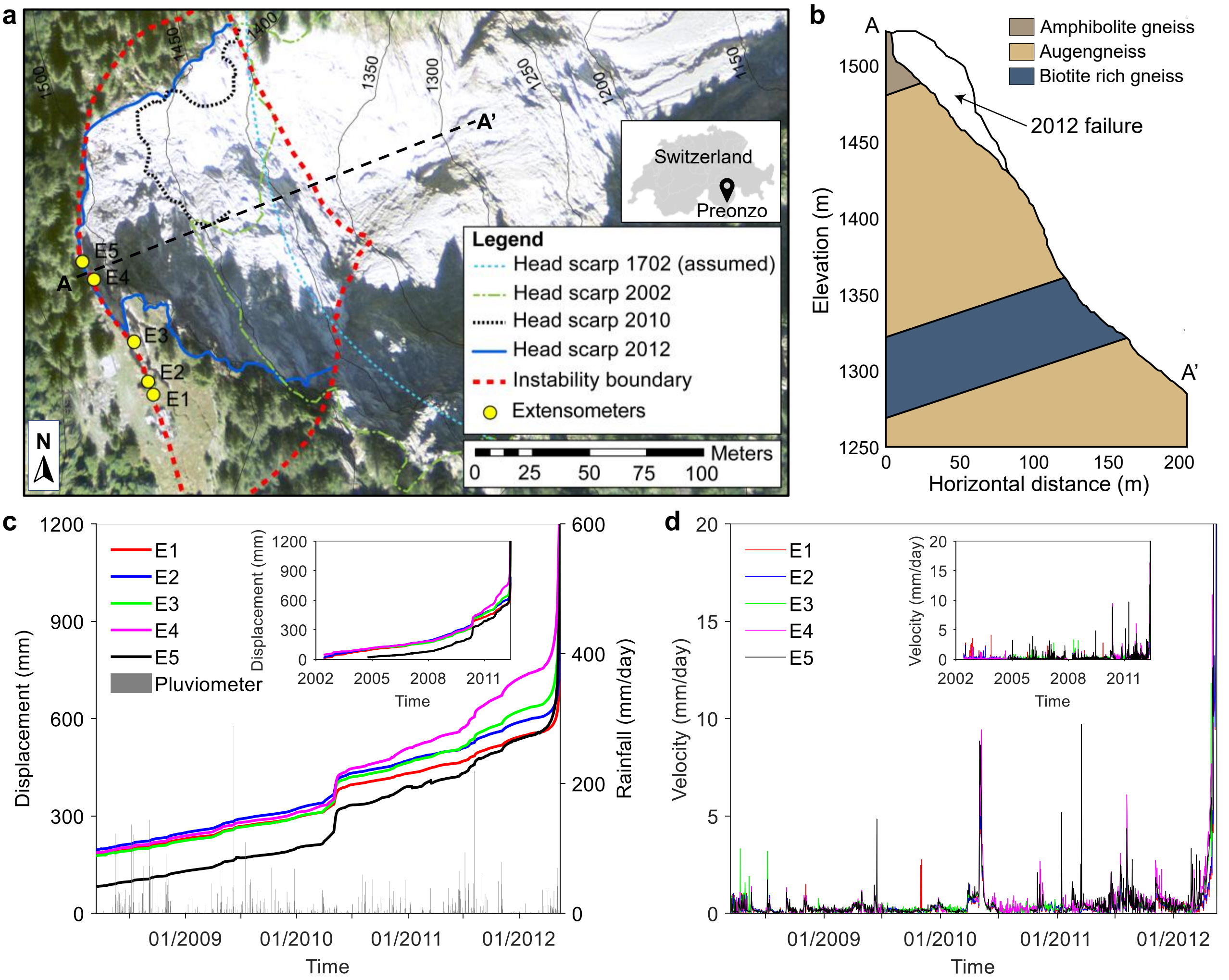

**Fig. 4 Preonzo landslide, Switzerland. a** Overview of the landslide site with the locations of five extensometers E1-E5, the boundary of this instability complex, the headscarps of historical failure events, and the cross-section line A-A' indicated. **b** Cross-section A-A' of the slope with the lithological units and the topographies (before and after the 2012 failure event) depicted. **c** Monitoring data of slope displacements by the five extensometers and recorded data of daily rainfall amount by a pluviometer installed at the slope. **d** Slope daily velocity time series derived from the displacement data.

**Application to the Preonzo landslide, Switzerland.** We test our theory based on the long-term monitoring dataset of a rainfall-induced landslide above the village of Preonzo, Switzerland[44] (Fig. 4a), which exhibited significant episodic movements over many years prior to a catastrophic failure in 2012. This instability complex is situated on a steep slope (dipping about 60° towards the Riviera valley), which is mainly composed of amphibolite gneiss, augengneiss, and biotite-rich gneiss (Fig. 4b). The slope is characterized by multiple sets of fractures and contains a large number of rock blocks[45], which is compatible with our conceptual model (Fig. 2). This landslide has experienced multiple failures since the 18th century[45]: the collapse in February 1702 destroyed the ancient village of Preonzo; in May 2002, ~120,000 $m^3$ of rock from the southern sector was detached; in May 2010, ~20,000 $m^3$ of rock from the northern sector was released; in May 2012, ~210,000 $m^3$ of rock catastrophically failed (see the headscarps of these historical events in Fig. 4a). The 2012 failure occurred in a mixed kinematic mode, involving oblique/flexural toppling in the headscarp region and planar/wedge sliding along the basal rupture plane formed by preexisting fractures and new cracks[45].

To closely monitor this active landslide that posed a great threat to the industrial and transport infrastructures located directly at the toe of the slope[44], five high-precision extensometers E1-E5 (Fig. 4a) were instrumented to measure the opening of tension cracks in the headscarp area. From 2008, a pluviometer was installed to monitor the local precipitation conditions. Fig. 4c shows the time series of slope displacement measured by the five extensometers and of daily rainfall amount recorded by the pluviometer between 2008 and 2012 (see the inset for the displacement time series from 2002 and Supplementary Fig. 3 for the time series of rainfall intensity and cumulative rainfall amount). We further compute slope velocities on a daily basis from the displacement data (Fig. 4d). One can see that this landslide exhibited a step-like deformation behavior over time (Fig. 4c) with strong velocity intermittency (Fig. 4d), resulting from short-term episodic movements (showing repeated temporary acceleration-deceleration cycles) superimposed on long-term deformation trends (showing an overall deceleration from 2002 to 2006, followed by an overall acceleration until 2012; Supplementary Figs. 4 and 5).

All the four types of episodic landslide dynamics, namely endogenous/exogenous-subcritical/critical, can be found in the velocity time series (see Fig. 5 for typical examples from individual extensometers and see Supplementary Figs. 6-10 for all five extensometers). We fit the data of normalized velocities to a power law (see Methods) to estimate the exponent $p$ with its standard deviation derived from the confidence interval of the fit.

For the Type I exogenous-subcritical peak on 7 May 2010 (Fig. 5a; extensometer E1), the velocity relaxation beyond ~8 days after the peak is characterized by an exponent of $p = 1.40 \pm 0.07$ (exogenous-subcritical) (Fig. 5a, inset), whereas its early-time relaxation within ~8 days is associated with a much smaller exponent of $p = 0.47 \pm 0.11$ (exogenous-critical), as expected from the prediction by equation (3). All five extensometers exhibit a similar two-branch power law relaxation behavior with $p = 0.46 \pm 0.10$ for the early-time response and $p = 1.54 \pm 0.06$ for the late-time response (Supplementary Figs. 6a and 7). Around this peak accompanied by little precipitation (Fig. 5a), the slope has experienced a localized failure in its northern sector downhill from the tension cracks where the extensometers are installed (Fig. 4a).

For the Type II exogenous-critical peak on 9 August 2011 (Fig. 5b; extensometer E1), the post-peak velocity relaxation obeys a power law with $p = 0.55 \pm 0.02$ (exogenous-critical) (Fig. 5b, inset). Prior to this peak, a heavy rainstorm has occurred (Fig. 5b). All five extensometers have captured this peak followed by a power law relaxation with $p = 0.63 \pm 0.03$ (Supplementary Figs.

6a and 8).

In Fig. 5c (extensometer E5), we present a Type III endogenous-subcritical peak preceded by no rainfall event. This peak is surrounded by an essentially noisy stationary velocity trajectory with $p \approx 0$ (Fig. 5c, inset), whereas most extensometers do not capture this peak and only show random fluctuations (Supplementary Figs. 6c and 9).

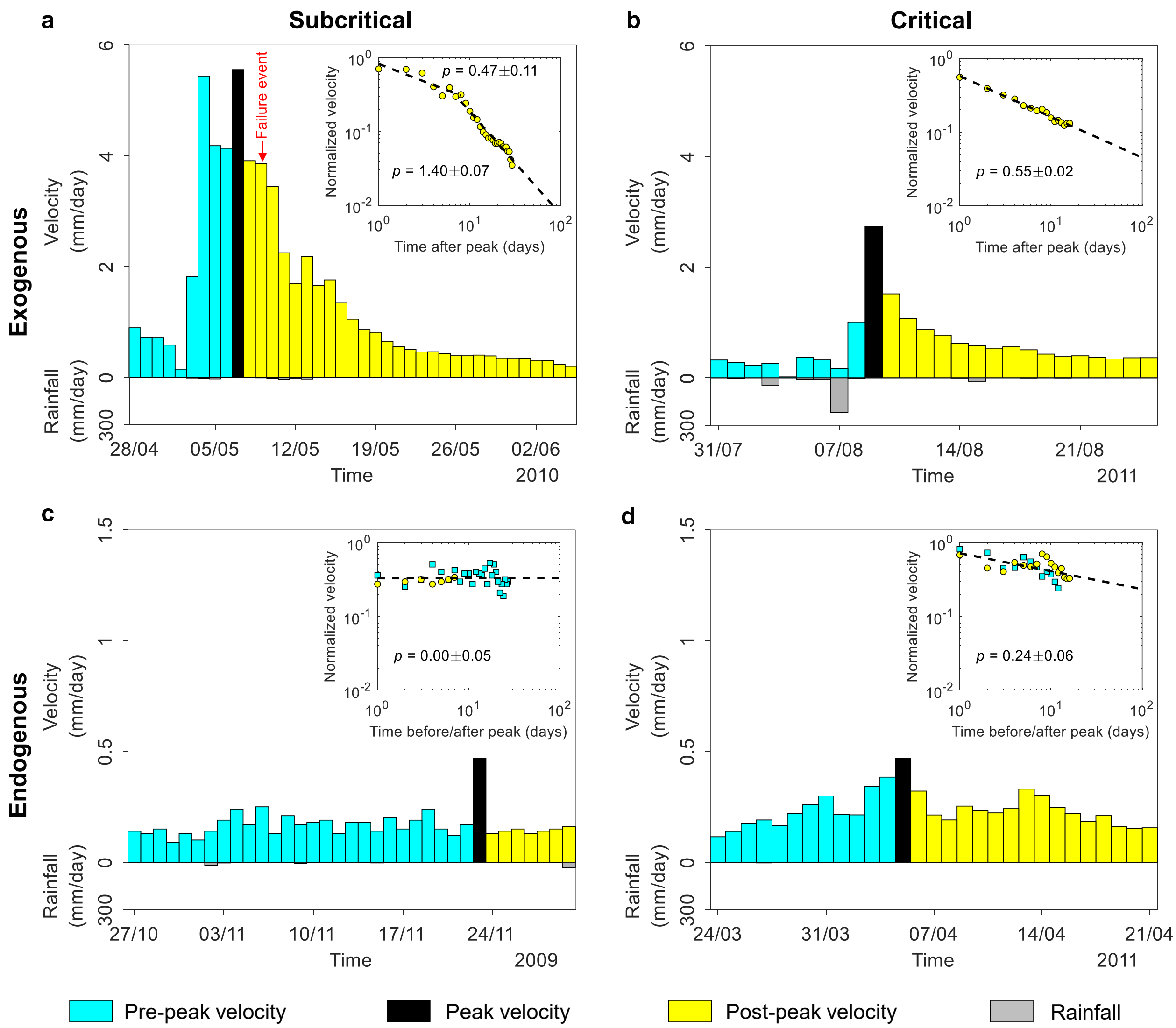

**Fig. 5 Four categories of episodic landslide dynamics found in the velocity time series of the Preonzo landslide.** **a** Type I, exogenous-subcritical; **b** Type II, exogenous-critical; **c** Type III, endogenous-subcritical; and **d** Type IV, endogenous-critical. Here, **a**, **b**, and **d** show velocity data from extensometer E1 and **c** shows velocity data from extensometer E5. The red arrow in **a** marks the timing of the local failure of a downslope northern sector of the slope on 9 May 2010. Insets show the post-peak relaxation of normalized velocity where dashed lines indicate the power law fitting (in the insets of **c** and **d**, pre-peak velocity data of those endogenous peaks are also included in the fitting).

Lastly, we show a Type IV endogenous-critical peak (Fig. 5d; extensometer E1), which is preceded by a progressively accelerating power law growth of velocity and followed by an approximately symmetrical power law relaxation, with a common exponent of $p = 0.24 \pm 0.06$. It seems that the majority of the five extensomers has captured such an approximately symmetrical precursory-recovery dynamics with a small power law exponent of $p = 0.21 \pm 0.04$ (Supplementary Fig. 6d), although the timing of the peaks recorded by individual extensometers is not fully synchronized (Supplementary Fig. 10). One can notice that the time-dependent signatures of endogenous peaks are less apparent compared to exogenous ones, as expected from the small exponent values.

Interpreting these results in light of equations (3) and (5)-(7), the obtained power laws for these different peak types have their different exponents $1+\vartheta$, $1-\vartheta$, 0, and $1-2\vartheta$ explained by a single parameter $\vartheta \approx 0.45 \pm 0.10$, providing strong support for our theory.

We develop an automated detection algorithm to identify velocity peaks with well-defined precursory/recovery characteristics from the 10-year long-term monitoring dataset of the Preonzo landslide (see Methods). In Fig. 6a, we show the histogram of their power law exponents $p$, which cluster into two distinct groups, one with a median at $p \approx 0.59$ and the other with a median at $p \approx 1.52$. This result is compatible with our theoretical prediction based on $\vartheta \approx 0.45 \pm 0.10$, yielding $p \approx 1.45 \pm 0.10$ for Type I peaks and $p \approx 0.55 \pm 0.10$ for Type II peaks. It seems that Type III and IV peaks (with $p \approx 0$ and $0.1 \pm 0.20$, respectively) are absent in Fig. 6a. This is because they usually have small magnitudes and considerably fluctuating post-peak responses (Fig. 5c-d and Supplementary Fig. 6c-d), making it difficult for them to pass the peak detection criteria (see Methods). We then compute the ensemble average of the relaxation behavior for the two exogenous peak types (Fig. 6b), with the fitted power laws consistent with the existence of a single parameter $\vartheta \approx 0.45 \pm 0.10$. This is particularly revealed by the equality (within statistical fluctuations) of the exponent of Type I peaks for $t - t_c < t^*$ $(p = 0.52 \pm 0.12)$ and of the exponent of Type II peaks $(p = 0.55 \pm 0.03)$.

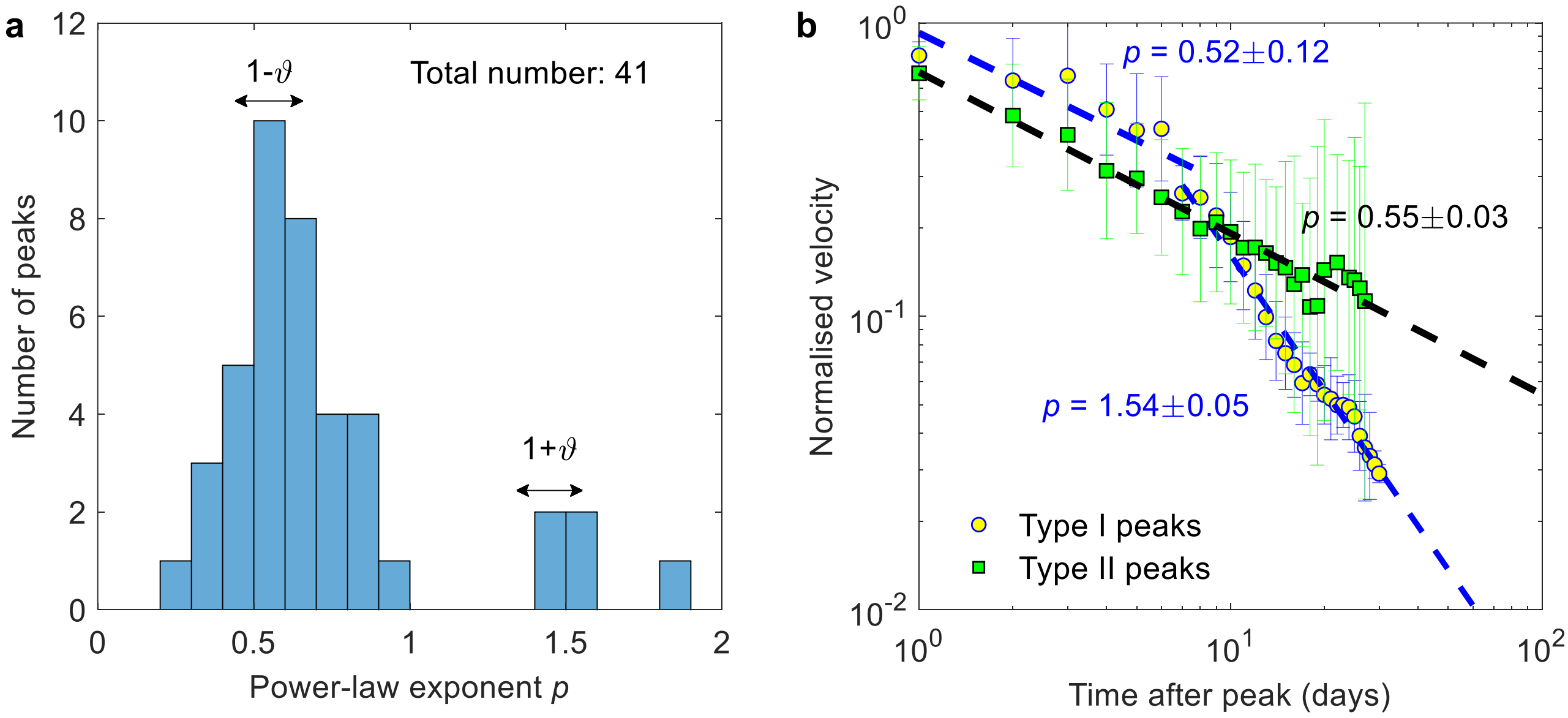

**Fig. 6 Post-peak relaxation properties associated with detected peaks in the velocity time series.** **a** Histogram of power law exponents $p$ for post-peak velocity relaxation; the double arrows indicate the value ranges of $p = 1-\vartheta$ (Type I peaks) and $p = 1+\vartheta$ (Type II peaks), with $\vartheta \approx 0.45 \pm 0.1$. **b** Ensemble averaged velocity relaxation behavior for Type I and II peaks; error bars indicate the standard deviation associated with the ensemble average.

## Discussion

We have presented a novel endo-exo theoretical framework to quantitatively classify episodic landslide movements into four fundamental types of distinct precursory/recovery signatures but related by a single common parameter $\vartheta$. All four power law regimes of episodic dynamics have been observed in the Preonzo landslide with $\vartheta \approx 0.45 \pm 0.10$ (see Figs. 5 and 6). While the $\vartheta$ value here is different from the value of $\vartheta = 0$ predicted by the fiber bundle model assuming a load shearing and stress transfer mechanism[33,46,47], a similar value of $\vartheta \approx 0.5$ has also been previously reported for shallow earthquakes[48], which can be explained by the first-passage problem associated with an underlying random walk of stress fluctuation[49]. Here, we build on this random stress fluctuation mechanism in the context of landslides. Consider that, at the moment when the movement of a mother mass stops, the stresses at its surrounding first-generation daughter masses are marginally lower than the critical threshold (i.e., strength). Let us assume the subsequent stress at each first-generation daughter mass fluctuates as a result of random increments stemming from various sources (small-scale damage/healing, poroelastic reorganization, small earthquake/rainfall events). The stress dynamics can thus be described by a Brownian random walk (Supplementary Fig. 10a). Then, the waiting time for a daughter mass to start moving is determined by the first time its stress exceeds the strength. This first-passage time is distributed according to a probability density function asymptotically converging to a power law[50] $\psi(t) \propto t^{-3/2}$, which yields $\vartheta = 0.5$ (Supplementary Fig. 10b). This result suggests that the mass movements of the Preonzo landslide tend to be more dominated by the intrinsic stress fluctuations driven by a continuous stochastic flow of small external perturbations (as embodied in the first-passage model), rather than by the load sharing and stress transfer processes (as assumed in the fiber bundle model[33,46,47]). The approximately normal diffusion behavior of stress fluctuations suggests that the Preonzo landslide is characterized by a relatively homogenous geological condition, which is consistent with previous field surveys indicating a generally uniform lithology[45] (with the minor heterogeneity causing $\vartheta$ slightly below 0.5 along with a small dispersion). This first-passage model also provides a natural interpretation to the origin of parameter $c$ in equation (2): it is simply the time interval for random walkers to culminate in crossing the gap between the initial stress level and the critical threshold; the shorter the gap is, the smaller the $c$ value is.

Our findings indicate that most large velocity peaks of the Preonzo landslide are exogenous-critical, suggesting that the landslide's behavior in reaction to strong external events (e.g., heavy rainfalls) is primarily driven by cascades across multiple generations of mass movement triggers. Consequently, the collective response of the mass system as a whole is slower and more sustained with an exponent $1 - \vartheta$, compared to the faster individual mass responses with an exponent of $1 + \vartheta$. In addition, we have documented a unique exogenous-subcritical type of episodic landslide dynamics, which is related to the local failure of a downslope sector[44,45] on 9 May 2010. Before showing a rapid exogenous-subcritical relaxation characterized by the large exponent $1 + \vartheta$, the landslide has actually experienced ~8 days of relatively slower exogenous-critical relaxation with the small exponent $1 - \vartheta$ (Figs. 5a and Supplementary Fig. 6a). Substituting this characteristic time $t^* \approx 8$ days together with $\vartheta \approx 0.45$ into equation (9) (see Methods) and the estimate $c \approx 1$ day, we obtain $n \approx 0.61$. This comparatively low branching ratio $n$ is consistent with the fact that this local failure did not cascade into a system-sized collapse. In our dataset, we also observe the presence of endogenous-critical dynamics (Fig. 5d). However, they are usually associated with small-magnitude peaks and weak time-dependence (governed by a small relaxation exponent of $1 - 2\vartheta \approx 0.1$), making them sometimes difficult to be discriminated from the endogenous-subcritical dynamics driven by random fluctuations. These results confirm the possibility of distinguishing endo/exo origins of episodic landslide movements by using our framework (see Supplementary Note 2 for more discussions).

We construct the time history of the branching ratio $n$ for the Preonzo landslide (see Supplementary Note 3 and Supplementary Fig. 11b). One can see that, before 2010, $n$ had an overall decline moving away from the critical state; around the 2010 local failure event, $n$ experienced a sudden jump moving closer to unity and eventually exceeded it when approaching the 2012 failure. Although the 2010 local failure did not cascade into a global rupture (because $n < 1$), the generated shock brought the system closer to criticality, promoting exogenous velocity peaks (mostly exogenous-critical peaks) (Supplementary Fig. 12) and facilitating the final collapse. Compared to the long-term deformation trends (Supplementary Figs. 4-5), we interpret that the Preonzo landslide transitioned from healing-dominated condition before 2006 to damage-dominated condition after 2010, driven by gravity-induced creep and precipitation-induced degradation.

Our results indicate that this landslide operated in the subcritical regime ($n < 1$) for many years and only approached the critical regime ($n \simeq 1$) for 1-2 years prior to the final collapse in 2012. This is consistent with the recent finding pointing out that fault systems also remain most of the time far from criticality[51]. However, this is in contradiction with the conventional concept of self-organized criticality[52], which holds that a landslide is permanently evolving at a critical state and all the events are generated by the same underlying process, making the prediction of large events impossible. Our new paradigm provides a more complete picture of landslide evolution, implying that characteristic signatures may emerge during the regime transitions to announce imminent catastrophic events (see below).

Up to now, we have mainly focused on the "endo-exo" regime where the landslide evolution is characterized by numerous accelerating-decelerating episodes driven by the interplay of exogenous perturbation and endogenous maturation. We now explore the last fundamental question raised in the Introduction section: how do episodic landslide movements relate to the final catastrophic failure? As the landslide progressively weakens, it could transition into the supercritical regime. This critical transition is found to be often endogenously driven in different natural and social systems[25], which rationalizes why many rainfall-induced landslides catastrophically fail in the absence of exceptional precipitation events[53]. If the supercritical regime is dominated by positive feedbacks with the slope acceleration behavior $\dot{v}(t) \propto v(t)^m$ characterized by $m > 1$, the system will exhibit a finite-time singularity and thus a catastrophic failure[54–56]. Note that a larger $m$ value indicates a stronger positive feedback effect. The solution of $v(t)$ also has the form of equation (3), but with exponent $p = 1/(m-1)$.

We fit the velocity time series of the Preonzo landslide prior to its major collapse on 15 May 2012 (Fig. 7a) to equation (3). We find it necessary to consider two power law branches, one with $p \approx 1.97$ ($m \approx 1.51$) for the early stage and the second one with $p \approx 0.49$ ($m \approx 3.04$) for the late stage (Fig. 7b). This suggests that the system is indeed dominated by positive feedbacks which seem to strengthen close to the final collapse. Our previous work showed that these late-stage large velocities are "dragon-kings"[55] (Fig. 7c) — a double metaphor for an event of a predominant impact/size like a "king" and a unique origin like a "dragon"[57]. They show statistically different properties compared to other smaller velocities that follow an inverse gamma distribution characterized by a power law tail exponent $\beta$ (see Fig. 7c and Supplementary Note 5). These dragon-kings can be detected as outliers via statistical tests[55] (see Supplementary Note 6 and Supplementary Figs. 13-24). This break in power law scaling thus marks the transition of the system from the self-organized criticality or "black-swan" regime[58] (where a catastrophic failure is unpredictable) to the dragon-king regime[55,57] (where a catastrophic failure becomes predictable) (Fig. 7b). Such a two-branch time-to-failure power law behavior has also been observed in the ground deformational response prior to the volcanic eruption at Mount St Helens[59]. We infer that the black-swan velocities (with $1 < m < 2$) result from the nucleation and propagation of approximately noninteracting cracks[60–62], while the dragon-king velocities (with $m > 2$) arise from crack interaction and coalescence leading to the formation of a global rupture plane[54]. This suggests that black-swans and dragon-kings are generated by distinct mechanisms, refining the existing theories[54–57]. Interestingly, when the Preonzo landslide entered the dragon-king regime, it once experienced a temporary deceleration during 7-11 May 2012 just before the final collapse. Such a precursory quiescence is consistent with the theoretical prediction for the supercritical regime[16] with $n > 1$ and $\vartheta > 0$. Substituting $t^* \approx 4$ days (Supplementary Fig. 15) and $\vartheta \approx 0.45$ into equation (9) (see Methods), which also holds for the supercritical regime[16], we obtain $n \approx 7.5$ (consistent with the estimated $n$ history in Fig. 7d), indicating an intense explosive branching process. Similar precursory quiescence phenomena have been observed prior to some great earthquakes[63] and may be attributed to dilatancy hardening[64], where pore pressure temporarily drops due to the pronounced dilation effect near the critical crack density[65]. This dilatancy may result from crack coalescence in rock bridges and/or frictional sliding along rough fractures, compatible with the field observation that the basal rupture plane is formed by preexisting fractures and new cracks[45].

Interestingly, the $\beta$-value of the slope velocity distribution progressively drops from $1.82$ to $1.62$ over 1-2 months prior to the final collapse (Fig. 7d), characterizing the system's subcritical/critical-to-supercritical regime transition (Supplementary Note 3). This indicates an increased occurrence of moderate to large velocities as the slope experiences the critical transition from the endo-exo regime (dominated by small velocities) to the black-swan regime (with progressively larger velocities growing in a self-similar manner as small velocities), and further to the dragon-king regime (dominated by large velocities that statistically and mechanistically differ from small velocities)[55]. The $\beta$-value decline prior to catastrophic landslides is similar to the observed $b$-value decline prior to great earthquakes[63,66], which is possibly due to increased differential stresses on rock bridges[67] and fault patches[68]. It also finds a natural explanation in the context of cascading triggered events described by self-excited conditional point processes[69]. This observation points to the possibility to forecast catastrophic landslides by monitoring the decline of the $\beta$-value (or the increase of the $n$-value) that may occur over a long term (say 1-2 months or even longer; Fig. 7d and Supplementary Fig. 11), in addition to detecting the emergence of dragon-kings[55] that may appear over a short term (say 1-2 weeks; Fig. 7c and Supplementary Figs. 13-14).

The system response in the supercritical regime has an entangled endo-exo origin due to the repetitive interactions and positive feedbacks[54,57], which rationalizes the ambiguous cause-effect relationship between the last rainfall event in early May 2012 and the final collapse about 10 days after[44] (Fig. 7a). In early May, this landslide was deep into the supercritical regime (with $n \gg 1$; Fig. 7d) and thus characterized by a high degree of endogeneity, making it inherently fragile and sensitive to external perturbations. The last rainfall then served as the "the straw that broke the camel's back" through noise amplification driven by endogenous positive feedbacks[56] (see Supplementary Note 2 for more discussions).

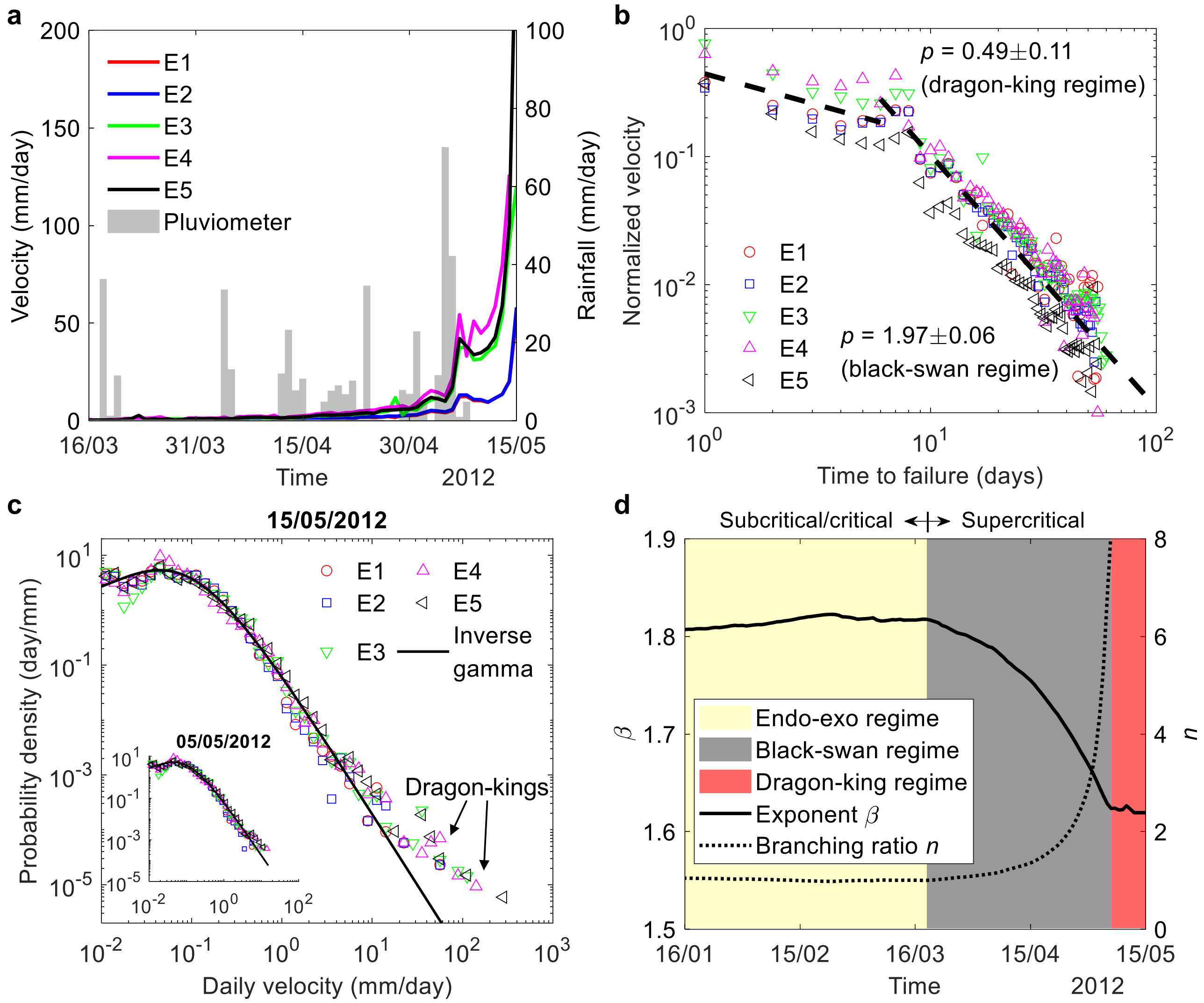

**Fig. 7 Evolution of the Preonzo landslide prior to its major collapse on 15 May 2012. a** Time series of the slope velocity measured by the five extensometers E1-E5 as well as rainfall intensity data recorded by the pluviometer. **b** Variation of normalized velocity prior to the catastrophic failure as a function of time to the failure (time flows from right to left), which is fitted to a two-branch finite-time singularity power law based on equation (3) (indicated by the dashed line). **c** Probability density function of daily velocities available until 15 May 2012 (inset: 06 May 2012) fitted to the inverse gamma distribution, where some large velocities manifest themselves as dragon-kings (outliers) deviating from the inverse gamma. **d** Progressive decline of the $\beta$-value of the velocity probability distribution and increase of the branching ratio $n$, indicating a transition of the landslide from a subcritical/critical (endo-exo) regime characterized by episodic movements to a supercritical (black-swan and dragon-king) regime ending with a catastrophic failure.

Our novel conceptual framework points at the existence of a deep quantitative relationship between episodic landslide movements, external triggering events, and internal damage/faulting/healing processes within the landmass. This framework allows us to distinguish the endo/exo origins of episodic landslide movements and further unravel their underlying mechanisms as well as their relationship with catastrophic failures. Compared to other empirical/physical models developed for slope creep and failure[70], our parsimonious endo-exo framework offers an evolutionary perspective to quantitatively diagnose episodic landslide movements and forecast catastrophic failure events, by capturing underlying regime shifts and emergent characteristic properties. Its formulation effectively captures the fundamental endo-exo mechanisms while retaining the flexibility to be calibrated with empirical field data. Although real-world data is used for calibration, this does not assume the model is inherently correct. Instead, the model functions as a conceptual framework to interpret physical phenomena and understand the observed behavior by identifying the key mechanisms governing it.

The results and insights obtained in the current work are of significant value for landslide hazard prediction and mitigation, from both the conceptual and operational points of view. Based on the well-documented dataset of the Preonzo landslide, we have provided a thorough validation of this framework, which can be applied to many other landslides showing similar episodic movements[6–14]. The broad applicability of our endo-exo solution, i.e., equation (3), across various landslide sites and types is supported by our analysis in Fig. 1 and Supplementary Figs. 1 and 2. Note that the parameter $\vartheta$ for other landslide sites could differ from $0.5$, especially if a strong heterogeneity is present leading to anomalous stress diffusion. Our observation of the decrease of exponent $\beta$ before catastrophic failures is consistent across various landslide sites and types, including rockfalls, topples, rockslides, and soilslides (see Supplementary Figs. 26 and 27, and Supplementary Table 1). This metric can be easily calculated in real time by fitting slope velocity data to the inverse gamma function using least squares or maximum likelihood methods, and readily integrated into early warning systems. The main practical constraint is the need for time-series monitoring data over a sufficiently long period (e.g., several months to years) to robustly constrain the $\beta$-value from the probability density function of slope velocities. Finally, it is worth mentioning that the established endo-exo framework has far-reaching implications for predicting and mitigating various geohazards, including not only landslides, but also earthquakes, rockbursts, volcanic eruptions, and glacier breakoffs, which all exhibit similar episodic deformations and sometimes also show transitions into catastrophic failures.

## Methods

**Mean field solution of the epidemic cascade model of mass interactions.** Consider a system that is subjected to a strong external shock with the exogenous activation $V(t)$ taking the form of a delta function:

$$V(t) = V_{\mathrm{c}}\delta(t - t_{\mathrm{c}}), \tag{4}$$

where $V_{\mathrm{c}}$ is the amplitude of the impulse occurring at time $t_{\mathrm{c}}$ and $\delta(t - t_{\mathrm{c}})$ is the unit impulse. Here, we use $V(t)$ to represent the effect of a strong external event that causes a sudden velocity response in the system. The conversion from external forces to $V(t)$ could occur through changes of the system's internal state (e.g., decreasing the resisting strength via effective stress reduction or material property alteration) and/or changes of the system's boundary condition (e.g., increasing the driving stress via load application or stress redistribution).

By substituting equation (4) into the second integral of equation (1), we can obtain:

$$v(t) = V_{\mathrm{c}}\delta(t - t_{\mathrm{c}}) + n\int_{-\infty}^{t} V_{\mathrm{c}}\delta(\tau - t_{\mathrm{c}})\Psi(t - \tau)\, d\tau = nV_{\mathrm{c}}\Psi(t - t_{\mathrm{c}}), \text{ for } t > t_{\mathrm{c}}. \tag{5}$$

This expresses that the post-peak dynamics is fully controlled by $\Psi(t - t_{\mathrm{c}})$, which is the Green function of the first equality of equation (1)[15], solution of:

$$\Psi(t - t_{\mathrm{c}}) = V_{\mathrm{c}}\delta(t - t_{\mathrm{c}}) + n\int_{-\infty}^{t} \psi(t - \tau + t_{\mathrm{c}})\Psi(\tau - t_{\mathrm{c}})\, d\tau. \tag{6}$$

For the power law form of $\psi(t - \tau)$ as defined in equation (2) and considering the normalization constraint $\int_{\tau+c}^{+\infty} \psi(t-\tau)dt = 1$, we have:

$$\psi(t-\tau) = \vartheta c^{\vartheta}/(t-\tau)^{1+\vartheta}, \text{ with } 0 < \vartheta < 1. \tag{7}$$

Then, the solution of equation (6) for $\Psi(t-t_{\mathrm{c}})$ can be obtained[16] by taking the Laplace transform of equation (6), which gives the Laplace transform of $\Psi(t-t_{\mathrm{c}})$, and then taking its inverse Laplace transform to yield:

$$v(t) \propto \Psi(t-t_{\mathrm{c}}) \propto \begin{cases} 1/(t-t_{\mathrm{c}})^{1-\vartheta}, \text{ for } c < t-t_{\mathrm{c}} < t^{*}, \\ 1/(t-t_{\mathrm{c}})^{1+\vartheta}, \text{ for } t-t_{\mathrm{c}} > t^{*}, \end{cases} \tag{8}$$

where $t^{*}$ is a characteristic time[16] given by:

$$t^{*} = c\left[\frac{n\Gamma(1-\vartheta)}{|1-n|}\right]^{1/\vartheta}, \tag{9}$$

where $\Gamma(\cdot)$ is the gamma function. As $n \to 1$ (critical regime), $t^{*} \to +\infty$, so that the early-time response prevails $(t-t_{\mathrm{c}} < t^{*})$; if $0 < n < 1$ (subcritical regime), $t^{*}$ has a finite value and the system may manifest a coexistence of both early-time response ($t-t_{\mathrm{c}} < t^{*}$, where epidemic cascades thrive) and late-time response ($t-t_{\mathrm{c}} > t^{*}$, where cascades become exhausted).

In the absence of any strong external event, a peak in landslide velocity may spontaneously emerge due to the interplay of a continuous stochastic flow of small external perturbations (introducing "noises" into the system) and the amplifying impact of the epidemic cascades of endogenous interactions. The pre and post-peak average velocity trajectory, conditioned on the existence of an endogenous peak $v_{\mathrm{c}}$ at time $t_{\mathrm{c}}$, is given by[15]:

$$v(t|v(t_{\mathrm{c}}) = v_{\mathrm{c}}) \approx \frac{v_{\mathrm{c}}}{\mathrm{var}(v_{\mathrm{c}})}\mathrm{cov}(v(t), v_{\mathrm{c}}), \tag{10}$$

for both $t < t_{\mathrm{c}}$ and $t > t_{\mathrm{c}}$. Since

$$\mathrm{cov}(v(t), v_{\mathrm{c}}) = \int_{-\infty}^{\min(t,t_{\mathrm{c}})} \Psi(t-\tau)\,\Psi(t_{\mathrm{c}}-\tau)d\tau, \tag{11}$$

and $\Psi(t-t_{\mathrm{c}}) \propto 1/(t-t_{\mathrm{c}})^{1-\vartheta}$ for $c < |t-t_{\mathrm{c}}| < t^{*}$, the following solution can be finally obtained[15]:

$$v(t) \propto \int_{-\infty}^{\min(t,t_{\mathrm{c}})} \Psi(t-\tau)\,\Psi(t_{\mathrm{c}}-\tau)d\tau \propto 1/|t-t_{\mathrm{c}}|^{1-2\vartheta}, \text{ for } c < |t-t_{\mathrm{c}}| < t^{*}, \tag{12}$$

or equivalently for $n \to 1$ (critical regime). If $n < 1$ (subcritical regime), the system response is essentially a noise process largely driven by random fluctuations, described by:

$$v(t) \propto 1/|t-t_{\mathrm{c}}|^{0}, \text{ for } |t-t_{\mathrm{c}}| > t^{*}. \tag{13}$$

**Calculation of normalized velocities around a peak.** We compute normalized slope velocities $\tilde{v}(t)$ around a peak based on the following equation:

$$\tilde{v}(t) = (v(t) - v_{\mathrm{r}})/(v_{\mathrm{c}} - v_{\mathrm{r}}), \tag{14}$$

where the slope velocity $v(t)$ at time $t = t_{\mathrm{c}}$ reaches a peak value $v(t_{\mathrm{c}}) = v_{\mathrm{c}}$ and $v_{\mathrm{r}}$ is the residual velocity when the landslide system has fully recovered from external perturbations. However, the determination of this residual velocity for a rainfall-induced landslide (like the Preonzo landslide) is subject to uncertainties, because the landslide has very rare opportunities to completely recover from one rainfall event before the next one occurs. In this work, we estimate the residual velocity by first detecting troughs in the velocity time series. We qualify a trough in the velocity time series as a local minimum over a 20-day time window which is at least $k = 2.5$ times smaller than the 2-month average velocity. The time window sizes and the threshold value $k$ are chosen to give an effective and reasonable detection of peaks and troughs from the data (see Supplementary Fig. 17), but the results do not significantly change by varying these parameters (see Supplementary Figs. 18-24). We then define the residual velocity associated with a given peak as the minimum of the two nearest troughs (with one before the peak and one after the peak). Note that this residual velocity tends to vary over time reflecting the nonstationary characteristic of the landslide. Supplementary Fig. 25 shows the probability density function of calculated residual velocities (associated with the identified peaks in Supplementary Fig. 17), which have a mean of 0.008 mm/day. We have also tested other possible approaches of determining the residual velocity, e.g., based on the average of the 10 nearest troughs around a peak or based on the minimum/average of the troughs located between the former peak and the latter peak. No significant changes in the results are found.

**Power law calibration of velocity time series around a peak.** We fit the time series of normalized velocities $\tilde{v}(t)$ around a peak to the finite-time singularity power law function:

$$\tilde{v}(t) = A/|t - t_{\mathrm{c}}|^{p}, \tag{15}$$

where $t_{\mathrm{c}}$ is the critical time chosen as the time of the peak, $A$ is a constant, and $p$ is the power law exponent. To estimate $A$ and $p$, we use the method of least squares to minimize the sum of squared residuals:

$$s = \sum_{t_i} \varepsilon(t_i)^2, \tag{16}$$

with each residual calculated as:

$$\varepsilon(t_i) = \log\tilde{v}(t_i) - \log A + p\log|t_i - t_{\mathrm{c}}|. \tag{17}$$

We then set the partial derivatives $\partial s/\partial(\log A)$ and $\partial s/\partial p$ to be both zero, leading to solve a linear system of two equations with the two unknowns $A$ and $p$.

**Automated velocity peak detection.** We implement a peak detection algorithm to automatically extract slope velocity peaks together with their surrounding time series from the 10-year long-term monitoring dataset. We qualify a peak in the velocity time series as a local maximum over a 20-day time window which is at least $k = 2.5$ times larger than the average velocity over a 2-month time window. The time window sizes and threshold value $k$ are chosen to give an effective detection of good-quality peaks (Supplementary Fig. 17), but the results do not significantly change by varying these parameters (Supplementary Figs. 18-24). In addition, we request that each peak has at least 10 days of post-peak data before reaching the next peak. In total, our algorithm detects 330 peaks from the entire dataset recorded by five extensometers. We then fit the post-peak velocity data of each detected peak to a power law over a time window ranging from 10 to 30 days, with the "best" window chosen as the one giving the highest coefficient of determination $R^2$. For the results shown in Fig. 6, we only keep the peaks with $R^2 > 0.8$ to extract unambiguous post-peak response functions, leaving 41 peaks. Our results do not qualitatively change by varying the $k$ threshold from 1.5 to 3.5 and the $R^2$ threshold from 0.7 to 0.9 as well as the window sizes for peak detection (Supplementary Figs. 22-24), suggesting that our method and results are robust. Despite the limited number of Type I peaks in our results (Fig. 6), the estimated median of their exponents $p$ remains robust when varying the peak detection parameters (Supplementary Figs. 22-24).

**Data availability**

The slope displacement monitoring data of the Preonzo landslide are publicly available at the ETH Zurich Research Collection (https://doi.org/10.3929/ethz-b-000600495).

**Code availability**

The computer code is available upon reasonable request from the authors.

**Acknowledgement**

Q.L. is grateful for the support by the Swiss National Science Foundation (Grant No. 189882) and the National Natural Science Foundation of China (Grant No. 41961134032). D.S. acknowledges partial support from the National Natural Science Foundation of China (Grant No. U2039202, T2350710802), from the Shenzhen Science and Technology Innovation Commission (Grant No. GJHZ20210705141805017) and the Center for Computational Science and Engineering at the Southern University of Science and Technology.

**Author contributions**

D.S. conceived the endo-exo framework; Q.L. and D.S. designed the research; Q.L. conducted the research; Q.L. and D.S. analyzed the results; Q.L. wrote the manuscript; Q.L. and D.S. reviewed and edited the manuscript.

**Competing interests**

The authors declare no competing interests.

**Additional information**

**Correspondence and requests for materials** should be addressed to Q.L.

## Supplementary Information for:
## Exo-endo framework for a unifying classification of episodic landslide movements

Qinghua Lei[1,*], Didier Sornette[2]

[1]*Department of Earth Sciences, Uppsala University, Uppsala, Sweden*

[2]*Institute of Risk Analysis, Prediction and Management, Academy for Advanced Interdisciplinary Studies, Southern University of Science and Technology, Shenzhen, China*

**This document includes:**

Supplementary Notes 1-5.
Supplementary Figs. 1-27.
Supplementary Table 1.

## Supplementary Note 1

We compile the monitoring data of 25 landslides, covering different landslide types including rockfalls, topples, rockslides, soilslides, and earthflows, recorded by different instruments (e.g., extensometers, reflectors, distometers, inclinometers, satellites, continuous GPS stations, and synthetic aperture radar) (see Supplementary Table 1 for more details about their locations, types, materials, volumes, and monitoring methods).

In Supplementary Fig. 1, we show a number of exogenous peaks observed in various landslides. Specifically, Supplementary Fig. 1a shows a rainfall-induced exogenous critical peak at the Veslemannen rockslide (Norway)[1]. Supplementary Fig. 1b shows an exogenous critical peak induced by a temperature rise at the Twain Harte granitic exfoliation dome (USA)[2]. Supplementary Fig. 1c shows a rainfall-induced exogenous critical peak at the La Saxe rockslide (Italy)[3]. Supplementary Fig. 1d shows a rainfall-induced exogenous critical peak at the Pomarico soilslide/earthflow (Italy)[4]. Supplementary Fig. 1e shows an earthquake-induced exogenous critical peak at the Maca soilslide (Peru)[5]. Supplementary Fig. 1f shows an earthquake-induced exogenous subcritical peak at the Tapgaon rockslide (Nepal)[6]. Supplementary Fig. 1g shows a rainfall-induced exogenous subcritical peak at the Yaoshan soilslide (PR China)[7]. Supplementary Fig. 1h shows a rainfall-induced exogeneous subcritical peak at the Hollin Hill rockslide[8].

In Supplementary Fig. 2, we further present a number of endogenous peaks observed in various landslides. Supplementary Fig. 2a shows an endogenous critical peak at the Veslemannen rockslide (Norway)[1]. Supplementary Fig. 2b shows an endogenous critical peak at the Twain Harte granitic exfoliation dome (USA)[2]. Supplementary Fig. 2c shows an endogenous critical peak at the Moosfluh rockslide (Switzerland)[9]. Supplementary Fig. 2d shows an endogenous critical peak at the Shuping soilslide (PR China)[10]. Supplementary Fig. 2e shows an endogenous critical peak at the Oak Ridge earthflow[11]. Supplementary Fig. 2f shows an endogenous critical peak (associated with a rockfall event) at the Ingelsberg rock slope (Austria)[12]. Supplementary Fig. 2g and 2h respectively shows an endogenous critical peak and an endogenous subcritical peak at the Séchilienne rockslide (France)[13].

These examples demonstrate the prevalence of endogenous and exogenous peaks in various types of landslides across diverse geological sites worldwide. While these examples were selected for illustration, additional cases are available in the dataset.

* Corresponding author: qinghua.lei@geo.uu.se

## Supplementary Note 2

In a system operating within the subcritical or critical regime, it is possible to differentiate between endogenous and exogenous origins of system responses. This distinction has been demonstrated across various contexts in the literature, as outlined in the introduction of the main text. This differentiation relies on the fact that a complex system responds differently to exogenous shocks compared to endogenous peaks, as demonstrated in our analytical solutions (see equations (8), (12), and (13) in Methods), as schematically illustrated in Fig. 3, and as further demonstrated in our application to the Preonzo landslide (see Fig. 5 and Supplementary Fig. 6). In this paper, we focus primarily on the subcritical/critical regime for the endo-exo classification. It is important to clarify that this differentiation between endogenous and exogenous origins refers to the initial triggering source—whether internal or external—rather than the dynamics of the system's response, which are always governed by endogenous processes.

It is important to emphasize that endogenous and exogenous factors are always interconnected, provided the system is not trivially purely exogenous (i.e., the branching ratio $n$ is not zero) and remains coupled to external influences—such as a continuous flow of noise fluctuations and sporadic large external events. First, let us consider an extreme case in which a system is purely exogenous ($n = 0$). In this scenario, external events would trigger only localized reactions within the system, resulting in isolated exogenous shocks, while the rest of the system remains entirely static without any fluctuations. Naturally, such an idealized system does not exist in reality. As long as $n > 0$, which is the general situation representing an internal organization of coupled subsystems, a large external event would have the tendency to trigger further events within the system, as we see in the exo-subcritical regime and exo-critical regime. For example, in the exo-critical regime (Type II), the initial shock is exo, and it is followed by a potentially large cascade of triggered events expressing the endogenous dynamics. On the other hand, in the endo-critical regime (Type IV), the system is driven, as usual, by a myriad of small external noise fluctuations which are renormalized into cascades of endogenously triggered events. These triggered events can create spontaneously growing activity, culminating in an endo peak. The dynamics of activities before and after the endo peak are approximately symmetric, in contrast with the very strong asymmetry of an exo-critical shock. Here, the physics and dynamics are actually the same as for the exo cases (given $n > 0$), namely the driving mechanism is always the self-excitation and triggering within the system in the presence of small fluctuations (“noise”) driving the system. In other words, our model and classification framework acknowledge the intrinsic coupling between exogenous events and endogenous triggering processes. This should not be confused with our effort to distinguish exogenous and endogenous origins, which refers to the initial triggering source of system responses (whether external or internal), rather than the underlying drive of these responses, which is consistently governed by self-excitation and triggering—processes that can be regarded as endogenous.

Moreover, an entangled endo-exo scenario arises in the supercritical regime, where the system becomes unstable and has a finite probability of experiencing exponential or super-exponential cascades of events. It is important to note that the driving mechanism remains the same as in the subcritical and critical regimes, with the system’s response still governed by self-excitation and triggering. This supercritical regime corresponds to a situation where different constituents of the system are strongly interacting with each other. Then, there is no clear cause-effect relationship in the system, which is of high endogeneity/fragility and very sensitive to exogeneous perturbations. The perturbation from an external event, even if minor, can be significantly amplified by endogenous positive feedbacks, explaining the ambiguous cause-effect relationship between the early May 2012 rainfall and the eventual collapse of the Preonzo slope roughly 10 days later. It is important to clarify that this does not imply a more dominant exogenous component; rather, it is the endogenous component that becomes more influential, with a branching ratio $n > 1$ driving an explosive branching process and intense cascade of disturbances. Thus, it is the endogenous processes that amplify the impact of an exogenous event, making the exogenous effect appear more prominent.

It is worth noting subtle differences in the endo/exo origin of system responses between the subcritical/critical regimes, where an endo-exo classification is feasible, and the supercritical regime, where such classification becomes challenging. For instance, in exo-critical landslide dynamics, a system response initiated by an exogenous trigger is endogenously amplified through cascades of mass interactions; however, the origin of this chain reaction is clearly the exogenous trigger. In other words, if there is no exo trigger, there is no shock and no associated recovery. This is different from an entangled endo-exo scenario in the supercritical regime. For the 2012 failure event of the Preonzo slope, if there were no rainfall in early May 2012, the slope might still fail eventually, but the rainfall indeed somehow made the failure to occur earlier. So, it is difficult to attribute the initial triggering source of this final failure to an endo or exo origin (but of course, the drive of the system's response is endogenous triggering processes).

## Supplementary Note 3

The results in the Main Text have been obtained by using the first-order moment (or average) of the landslide mass velocity, i.e., equation (1). We can improve the theory by accounting for the fact that the full process is stochastic with each mother mass motion potentially triggering a number of daughter mass motions according to the productivity law of mass movement triggering:

$$\rho(E) = \rho_0 (E/E_0)^a, \quad \text{(S1)}$$

defining the average number of daughter masses triggered by a mother mass of energy release $E \geq E_0$, where $E_0$ is an energy cutoff (i.e., a mass with energy release below $E_0$ does not trigger daughter masses), and $\rho_0$ and $a$ are positive constants.

Drawing parallels between landslides and earthquakes[5,11,14–18], we also postulate that the probability density function of daily energy release of a landslide follows a Gutenberg-Richter-type law as:

$$f(E(t)) \propto E(t)^{-(1+\mu)}, \quad \text{(S2)}$$

where $\mu$ is a positive exponent. Given $E(t) \propto v(t)^2$ and by applying the law of conservation of probability under a change of variable[38], we obtain the probability density function of daily velocities as:

$$f(v(t)) \propto v(t)^{-(1+2\mu)}. \quad \text{(S3)}$$

Since the probability distribution of the $v(t)$'s of the Preonzo landslide follows an inverse gamma distribution (with $\beta$ denoting its shape parameter; see Supplementary Note 4) characterized by a power law tail $f(v(t)) \propto v(t)^{-(1+\beta)}$, we therefore obtain:

$$\beta = 2\mu. \quad \text{(S4)}$$

Given $\mu = 2b/3$ and the stress-dependence of the seismic $b$-value[19], we may expect $b \approx 1.2 - 1.4$, $\mu \approx 0.80 - 0.93$, and therefore $\beta \approx 1.60 - 1.87$ for landslides (given the differential stress being[18] about a few MPa). This is consistent with the $\beta$ value we obtain in the current study for the Preonzo landslide in the supercritical regime (see Fig. 7d and Supplementary Fig. 11b) as well as the $\beta$ values of two other landslides during their crises obtained in our previous study[20].

The branching ratio $n$ is nothing but the average productivity and given by[21]:

$$n = \int_{\tau}^{+\infty} \int_{E_0}^{+\infty} f(E)\rho(E)\psi(t-\tau)dEdt. \tag{S5}$$

For $\vartheta > 0$ and $a < \mu$, given equation (7) in Methods, the double integral in equation (S5) leads to a finite value for $n$[21]:

$$n = \rho_0\mu/(\mu - a) = \rho_0\beta/(\beta - 2a). \tag{S6}$$

From equation (9) in Methods, we obtain, for the subcritical regime:

$$n = \left[1 + \Gamma(1-\vartheta)(c/t^*)^{\vartheta}\right]^{-1}. \tag{S7}$$

By combing equations (S6) and (S7), we obtain:

$$\rho_0 = (1 - 2a/\beta)\left[1 + \Gamma(1-\vartheta)(c/t^*)^{\vartheta}\right]^{-1}. \tag{S8}$$

For the Type I exogenous-subcritical peak observed at the Preonzo landslide (see Fig. 5a and Supplementary Fig. 6a), we have the characteristic time $t^* \approx 8$ days. According to the literature of earthquake aftershocks[22,23], $c$ may vary from days to minutes. Here, we estimate $c \approx 1$ day based on the power law decay trends of the Preonzo landslide around peaks, while a smaller $c$ (e.g., $0.1$ day) tends to overestimate $n$ (see Supplementary Fig. 11b). Furthermore, we postulate $a \approx 0.8$, which is the typical value for earthquakes[24] and proven to also hold for the Preonzo landslide (see below). Given $\vartheta \approx 0.45$, using equation (S7), we can obtain $n \approx 0.61$ for this exogenous-subcritical peak. Based on the profile maximum likelihood estimation[25], we find $\beta \approx 2.0$ during the relaxation stage of this peak (see Supplementary Fig. 11a). By using equation (S8), we further constrain $\rho_0 \approx 0.12$. Finally, based on the time series of the $\beta$ value (Supplementary Fig. 11a) determined by fitting the velocity data to the inverse gamma distribution (Supplementary Note 4), we obtain an estimation of the temporal evolution of $n$ (Supplementary Fig. 11b).

Note that the calculation of $n$ here is intended only as a first-order approximation, where local fluctuations are generally smoothed out in deriving $\beta$ from the probability density function of daily velocities. This explains the appearance of some exogeneous/endogenous-critical peaks when $n$ is not so close to $1$ (compare Supplementary Figs. 11b and 12). A more accurate estimation of $n$ may be achieved by calibrating the underlying self-exciting Hawkes point process to the velocity time series using the maximum likelihood estimation method[26], which we plan to explore in future work.

Since the total number of daughter masses triggered by a mother event of energy $E$ per day scales as[24] $\rho(E)f(E) \propto E^{a-\mu}$, we expect that the transition from the subcritical/critical regime (for which $a - \mu < 0$) to the supercritical regime (for which $a - \mu > 0$) is characterized by a shift from $a \lesssim \mu = \beta/2$ to $a > \mu = \beta/2$. Given $\beta$ progressively drops from $1.82$ to $1.62$ over 1-2 months prior to the final collapse (see Fig. 7d), we expect $a \approx 0.8 - 0.9$, which is consistent with our postulation above and comparable to the typical value of $a \approx 0.8$ for earthquakes[24]. This correspondence holds notwithstanding the fact that landslides happen in near-surface environments under low stress conditions, while earthquakes occur in deep subsurface regions subject to much higher stress levels. Our results demonstrating parallels between landslides and earthquakes provide additional supports for the fault mechanics perspective of landslide dynamics and failure[5,11,14–17]. It is also worth noting that some distinct catastrophic characteristics may exist between landslides and earthquakes. For example, the relationship between the $\vartheta$ value in the endo-exo regime and the $p$ value in the supercritical regime derived in the previous research in the context of earthquake aftershocks[27] does not explain our results for the Preonzo landslide. Further investigation is needed to better understand this problem.

## Supplementary Note 4

The probability density function of the three-parameter inverse gamma distribution is written as[25]:

$$f(v) = \frac{\alpha^{\beta}}{\Gamma(\beta)} \left(\frac{1}{v-\gamma}\right)^{\beta+1} \exp\left(-\frac{\alpha}{v-\gamma}\right), \tag{S9}$$

where $v$ is the slope velocity, $\alpha$ is a scale parameter, $\beta$ is a shape parameter equal to the exponent of the asymptotic power law tail for large $v$'s (according to the mathematical convention in the theory of Lévy stable laws[28]), $\gamma$ is a threshold velocity, and $\Gamma(\cdot)$ is the gamma function. The parameters need to meet the conditions of $\alpha > 0$, $\beta > 0$, and $\gamma < v$ for $f(v)$ to qualify as a normalized probability density function. The parameters $\alpha$, $\beta$, and $\gamma$ can be determined based on the profile maximum likelihood estimation method[25]. The inverse gamma distribution has an essential singularity at $v = \gamma$ with a rollover for $v$'s around the mode $\alpha/(\beta+1)+\gamma$, and a power law decay with a tail exponent $\beta$ for medium and large $v$ values, so that the tail of the inverse gamma is described by the power law[20]:

$$f(v) \approx \frac{\alpha^{\beta}}{\Gamma(\beta)} v^{-\beta-1}, \text{ for } v \gg \alpha+\gamma. \tag{S10}$$

## Supplementary Note 5

The detection of dragon-kings is achieved by performing a statistical test[25] on the slope velocity data $\{v_1 \leq \cdots \leq v_N\}$ ranked in ascending order. We define the null hypothesis $H_0$ that all the velocity data are drawn from the same population (i.e., according to the same inverse gamma distribution), whilst the alternative hypothesis $H_1$ states that the largest velocities follow a distribution different from that of their smaller siblings. We first identify outlier candidates based on the Anderson-Darling distance defined as[29]:

$$A^2(\xi) = -q - 2\sum_{i=1}^{q} \left\{\frac{2i}{2q+1} \ln[F(v_i)] + \left(1-\frac{2i}{2q+1}\right) \ln[1-F(v_i)]\right\}, \tag{S11}$$

where $F(\cdot)$ is the cumulative distribution function of the ranked velocity data $\{v_1 \leq \cdots \leq v_q \leq \xi; q \leq N\}$, such that the threshold velocity $\xi$ can be further estimated as[25]:

$$\hat{\xi} = \arg\min_{\xi} A^2(\xi), \tag{S12}$$

above which $l$ number of velocities are identified as outlier candidates (see Supplementary Fig. 13). We then conduct an outward sequential test by examining increasingly larger outlier candidates $v_j \in \{v_{N-l+1} \leq \cdots \leq v_N\}$ using a test statistic given by[25]:

$$T_{j,l}^{\text{outward}} = \frac{v_j}{\sum_{i=1}^{N-l} v_i}. \tag{S13}$$

For the remaining $r$ number of outliers $\{v_{N-r+1} \leq \cdots \leq v_N; 0 \leq r \leq l\}$, we conduct a block test using the following test statistic[25]:

$$T_r^{\text{block}} = \frac{\sum_{i=N-r+1}^{N} v_i}{\sum_{i=1}^{N-r} v_i}, \tag{S14}$$

and perform Monte Carlo simulation experiments (10,000 runs) to compute the probability value under $H_0$. A systematic decline of the probability value signifies the transition of the slope into the dragon-king regime (see Supplementary Fig. 14).

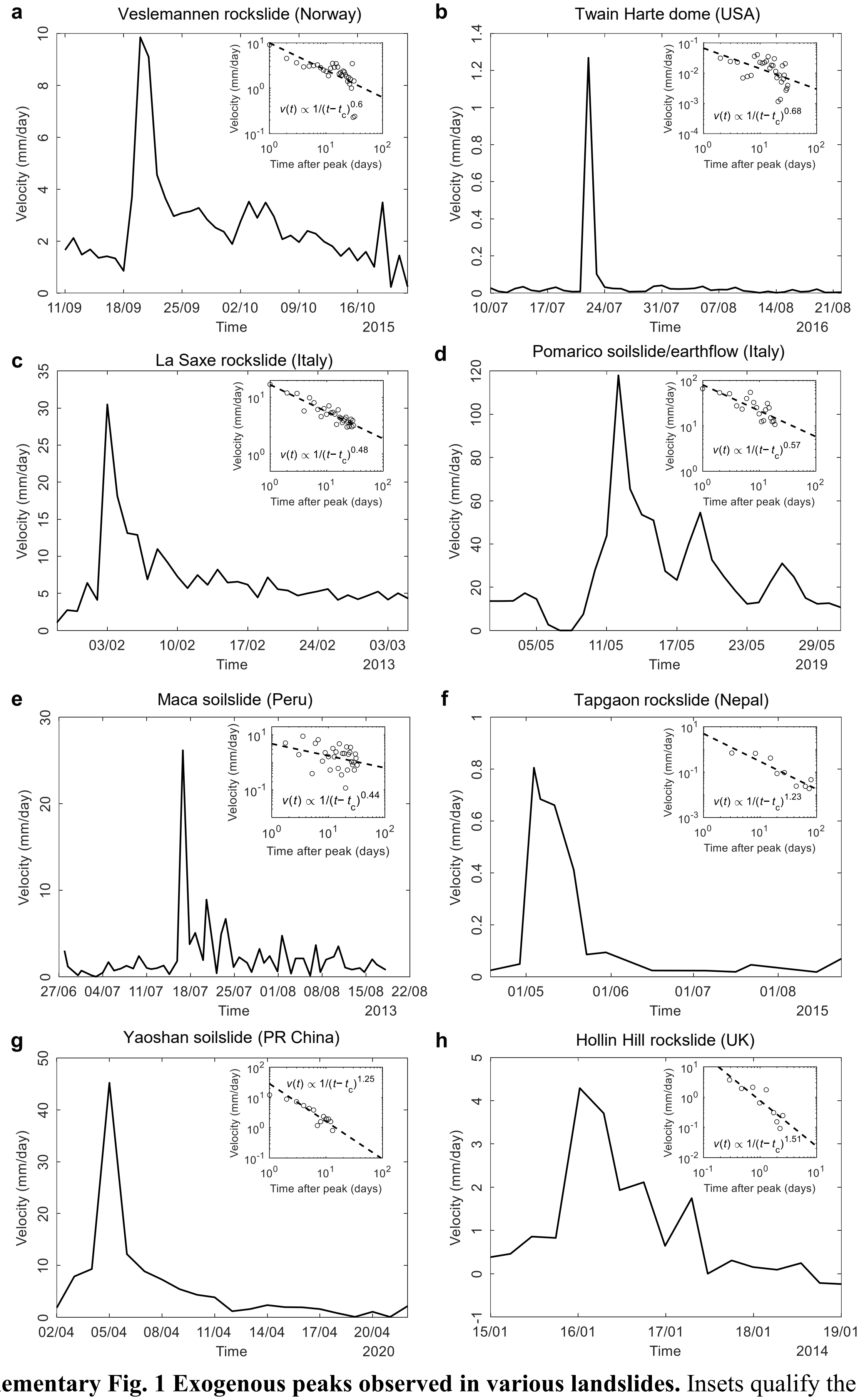

**Supplementary Fig. 1 Exogenous peaks observed in various landslides.** Insets qualify the power law singularity dynamics of the velocity $v(t)$ as a function of time after the peak at time $t_c$.

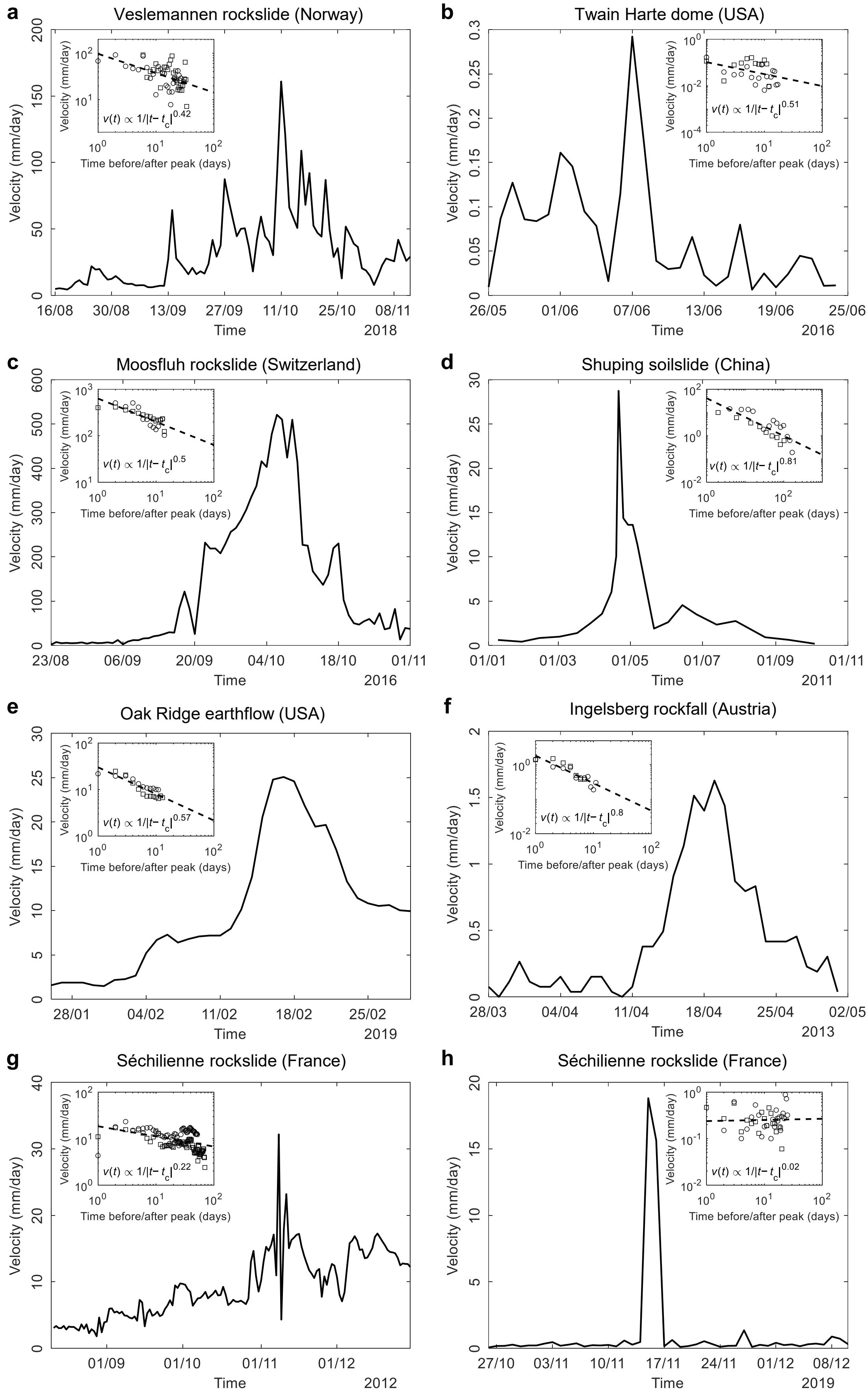

**Supplementary Fig. 2 Endogenous peaks observed in various landslides.** Insets qualify the power law singularity dynamics of the velocity $v(t)$ as a function of time before (squares) and after (circles) the peak at time $t_c$.

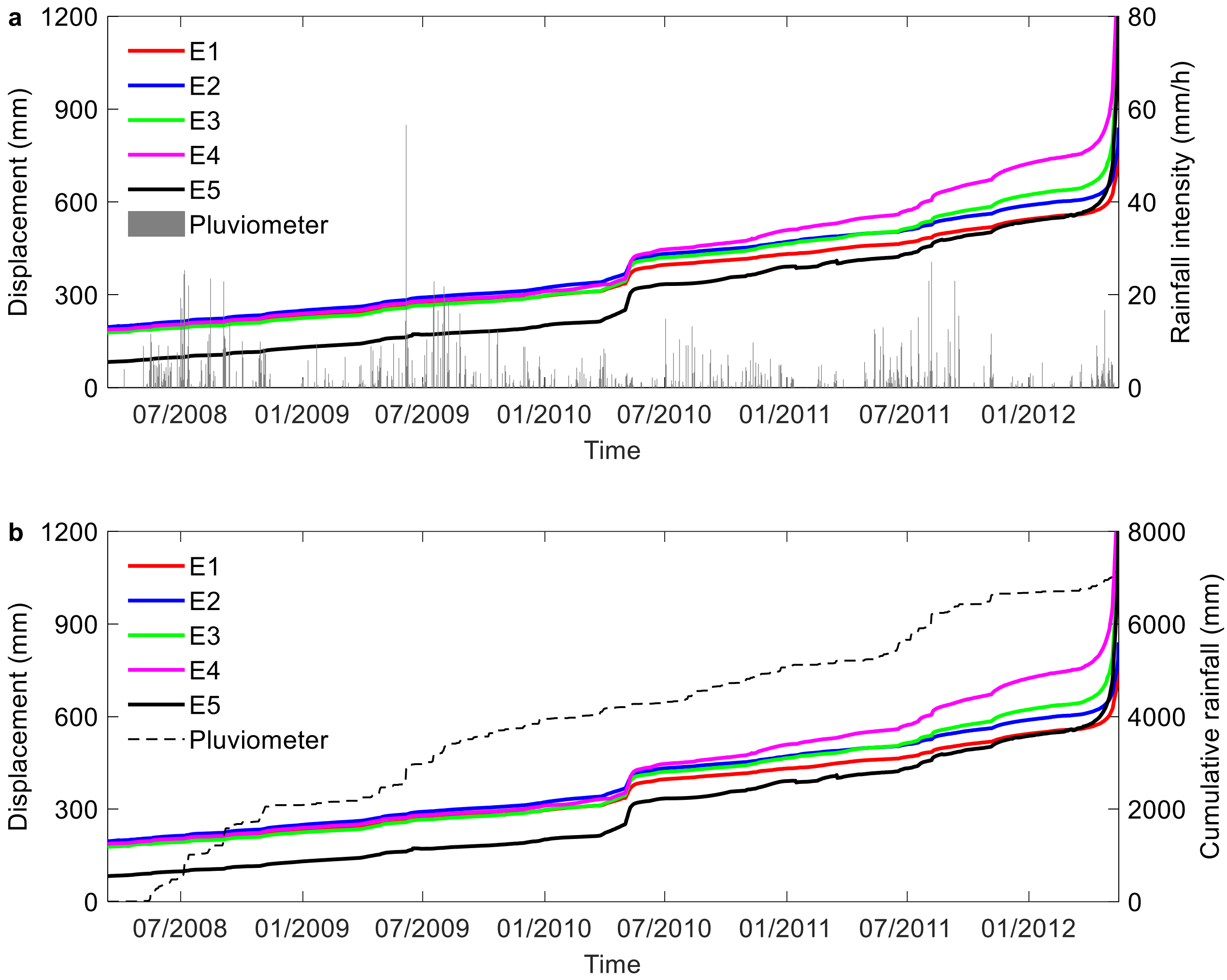

**Supplementary Fig. 3 Monitoring data of the Preonzo landslide, Switzerland.** Time series of slope displacements measured by five extensometers presented together with the data of **a** rainfall intensity and **b** cumulative rainfall amount recorded by a pluviometer installed at the Preonzo slope.

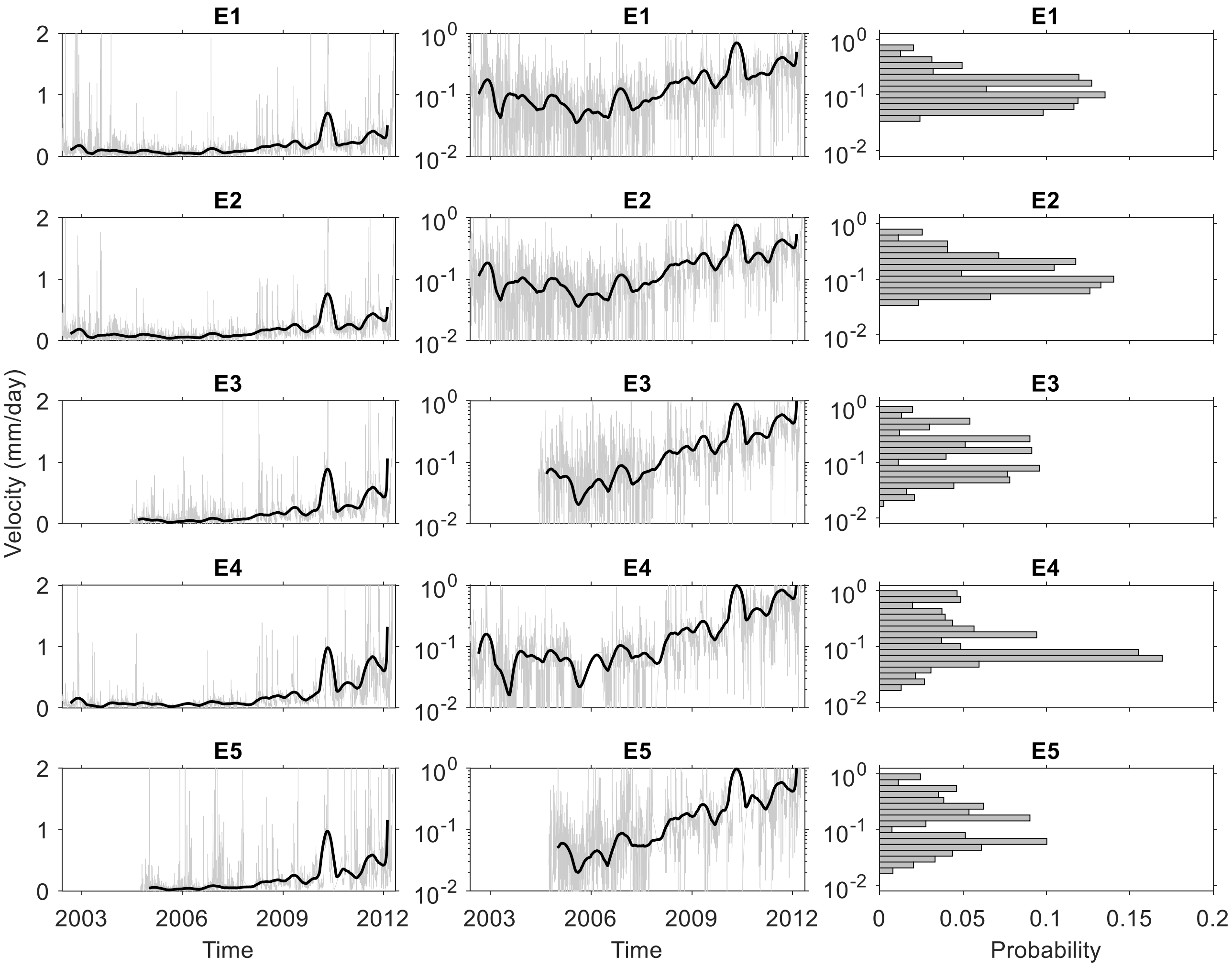

**Supplementary Fig. 4 Long-term velocity trends of the Preonzo landslide.** In the left and middle panels, velocities are plotted on the linear and logarithmic scales, respectively; the gray line gives the nonsmoothed velocity time series and the black line gives the smoothed velocity by the Savitzky-Golay filter with a polynomial order of 2 and a frame length of 6 months. The right panel gives histograms approximating the probability density function of long-term velocities.

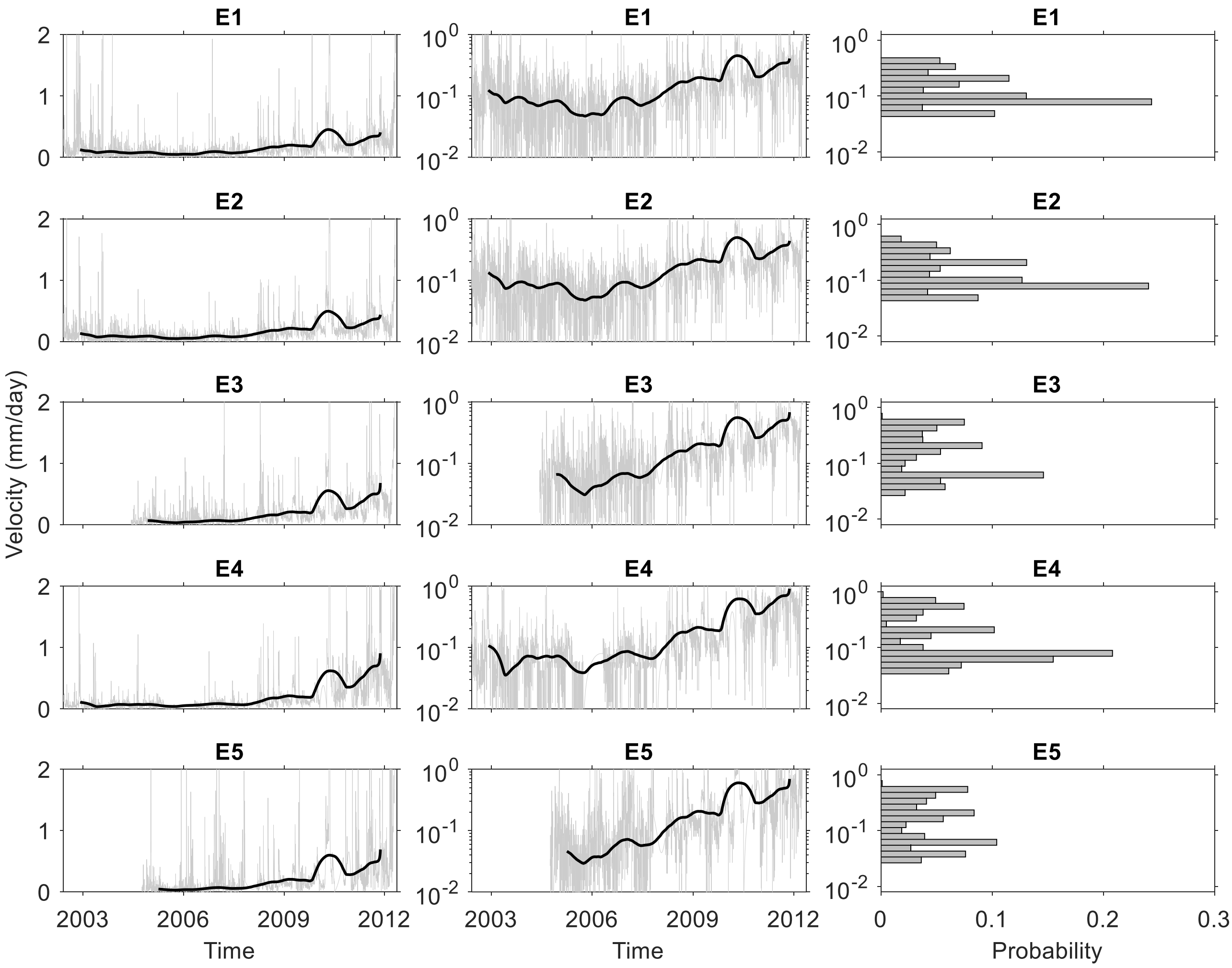

**Supplementary Fig. 5 Long-term velocity trends of the Preonzo landslide.** In the left and middle panels, velocities are plotted on the linear and logarithmic scales, respectively; the gray line gives the nonsmoothed velocity time series and the black line gives the smoothed velocity by the Savitzky-Golay filter with a polynomial order of 2 and a frame length of 1 year. The right panel gives histograms approximating the probability density function of long-term velocities.

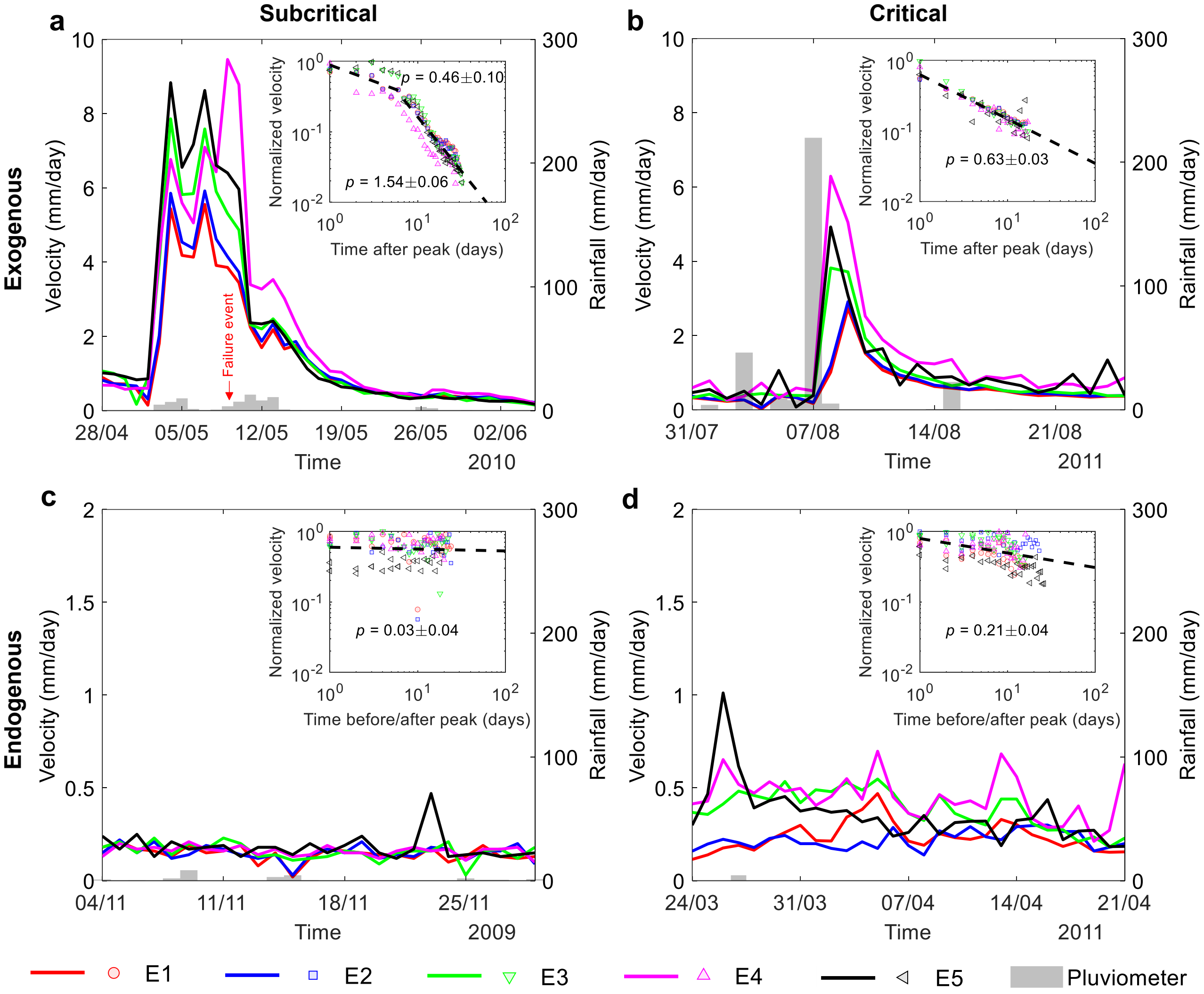

**Supplementary Fig. 6 Time series of daily slope velocity measured by the five extensometers E1-E5 and time series of rainfall data recorded by the pluviometer for different types of peaks (insets show the post-peak velocity relaxation).** **a** Type I, exogenous-subcritical; **b** Type II, exogenous-critical; **c** Type III, endogenous-subcritical; and **d** Type IV, endogenous-critical. The red arrow in **a** marks the timing of the local failure of a northern sector of the slope on 9 May 2010. For endogenous peaks as shown in **c** and **d**, pre-peak velocity data are also indicated (open markers) in addition to post-peak data (filled markers).

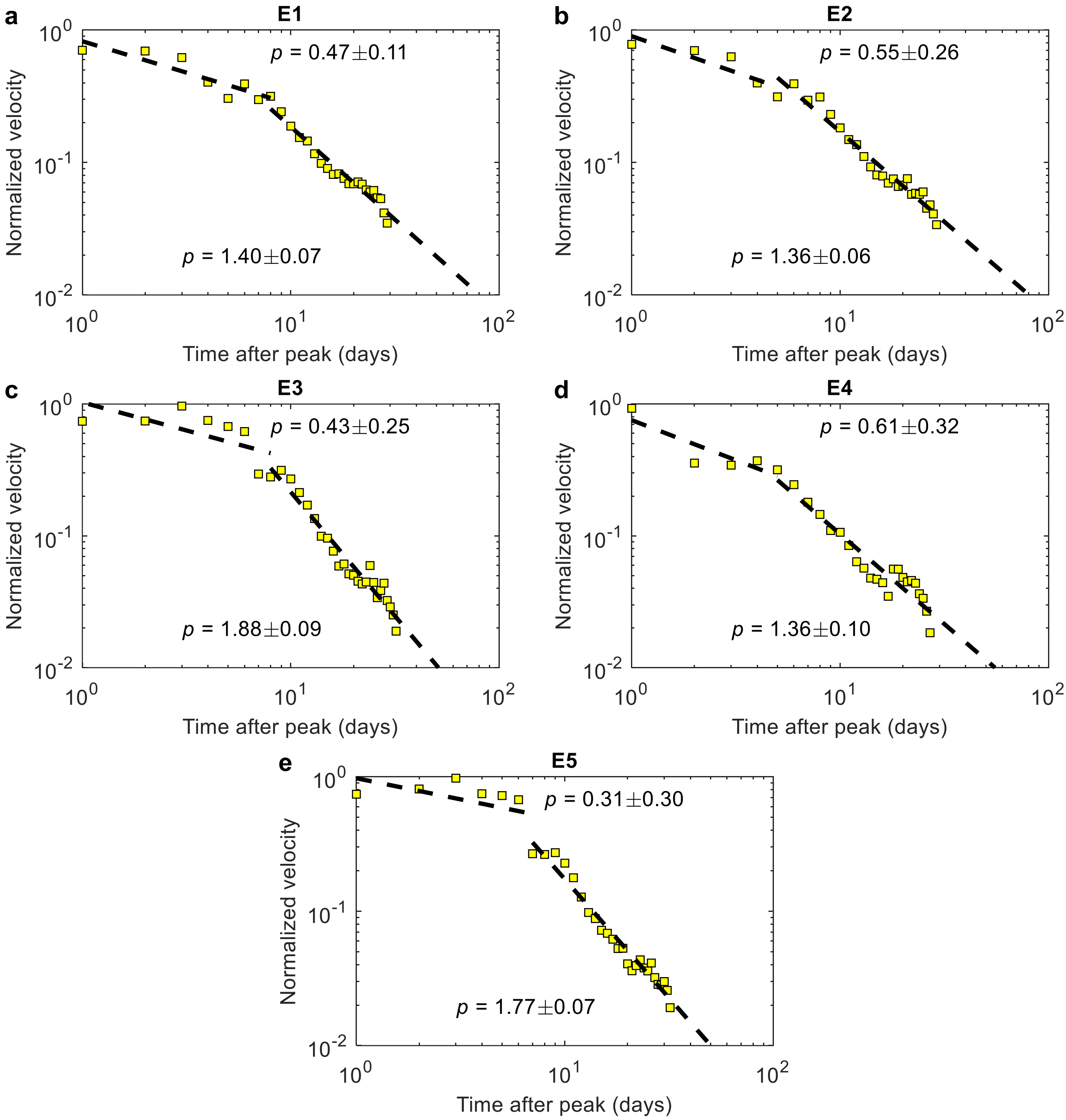

**Supplementary Fig. 7 Post-peak relaxation of Type I exogenous-subcritical peaks.** Variation of normalized velocity as a function of post-peak time for the five extensometers E1-E5.

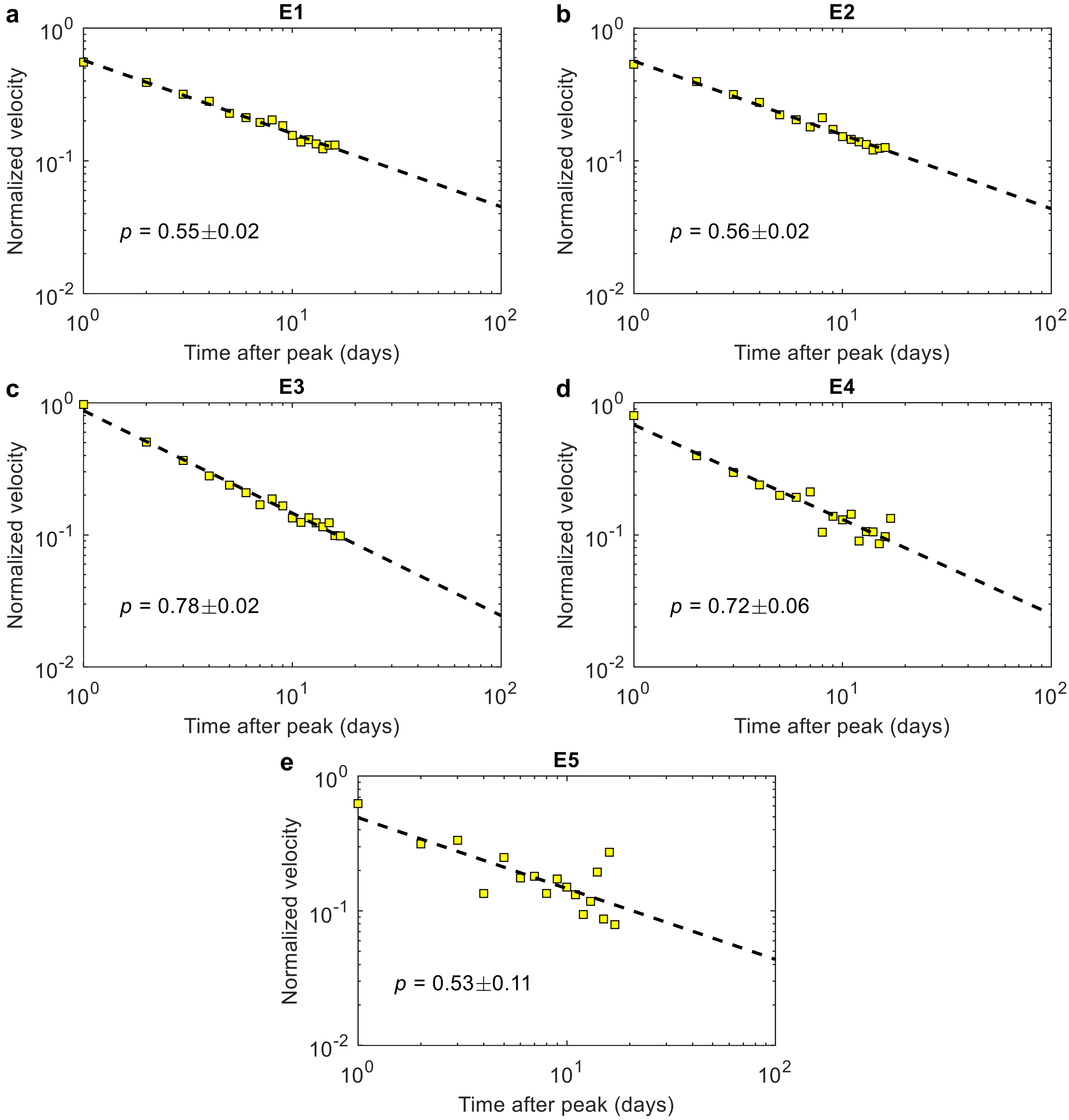

**Supplementary Fig. 8 Post-peak relaxation of Type II exogenous-critical peaks.** Variation of normalized velocity as a function of post-peak time for the five extensometers E1-E5.

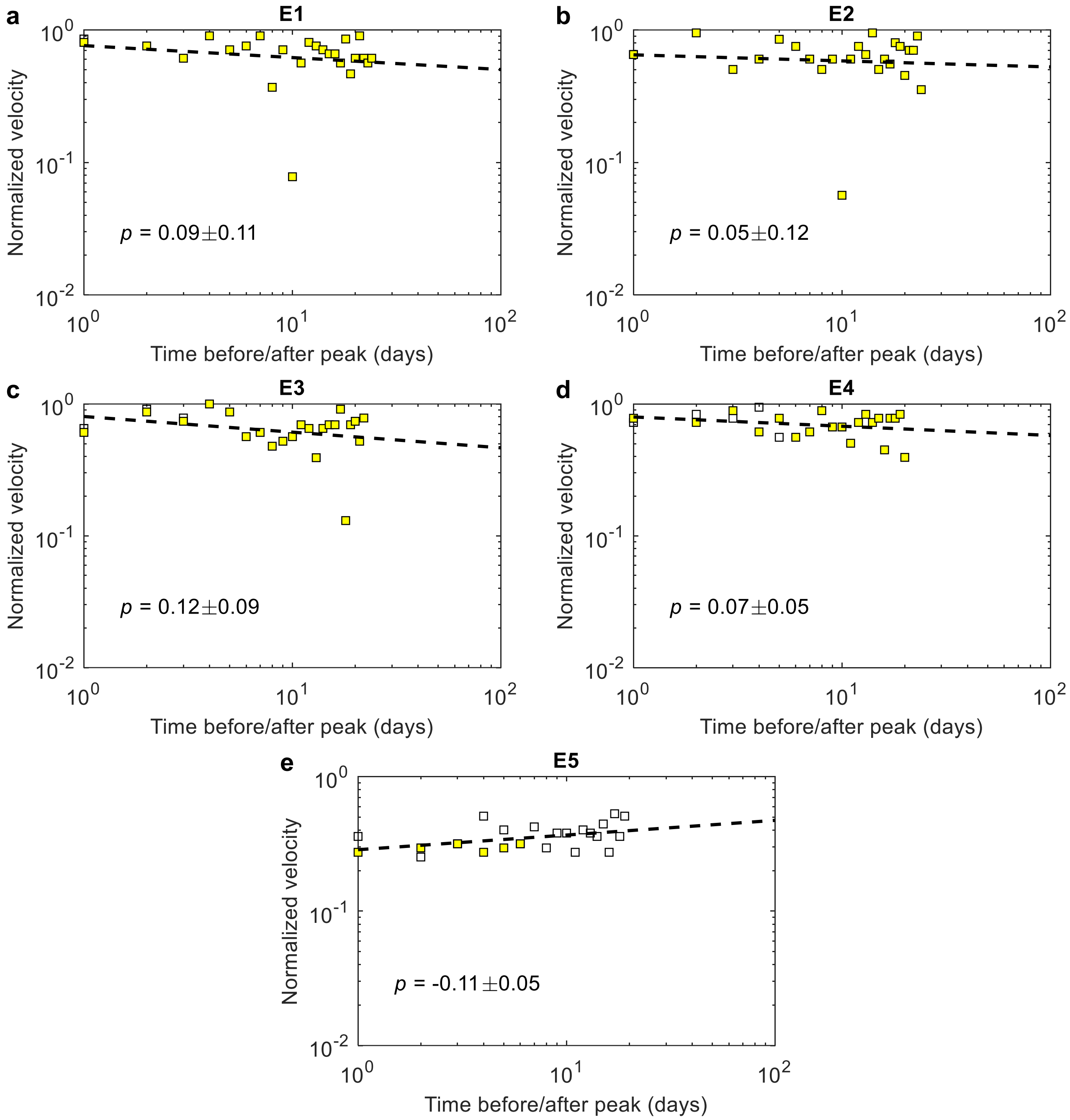

**Supplementary Fig. 9 Pre-peak (open symbols) acceleration and post-peak (colored symbols) relaxation of Type III endogenous-subcritical peaks.** Variation of normalized velocity as a function of pre/post-peak time for the five extensometers E1-E5.

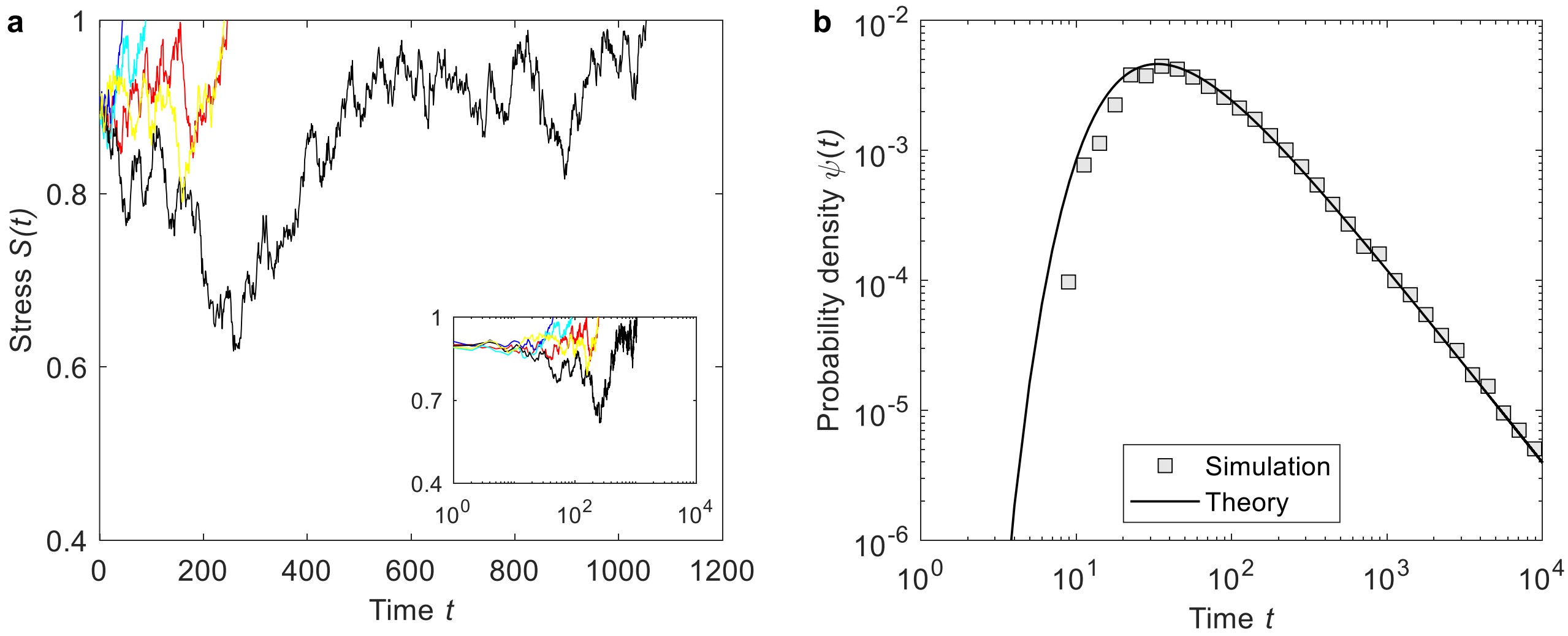

**Supplementary Fig. 10 Random walk simulation of stress fluctuations with a comparison to the theory of first-passage processes.** The random walk, with each step taking a unit time, is associated with a diffusion coefficient of $D = 1\times10^{-4}$. The waiting time corresponds to when the stress $S(t)$ with an initial value of $S_0 = 0.9$ for the first time exceeds the critical value of $S_c = 1.0$. **a** Stress history of 5 realizations (selected arbitrarily from 10000 realizations); inset shows the plot with time in log-log scale. **b** Probability density function of waiting times obtained from the random walk simulation with a comparison to the theoretical solution of the first-passage problem: $\psi(t) = (S_c - S_0)(2\pi D t^3)^{-1/2}\exp[-(S_c - S_0)^2/(2Dt)]$, which asymptotically converges to a power law $\psi(t) \propto t^{-3/2}$.

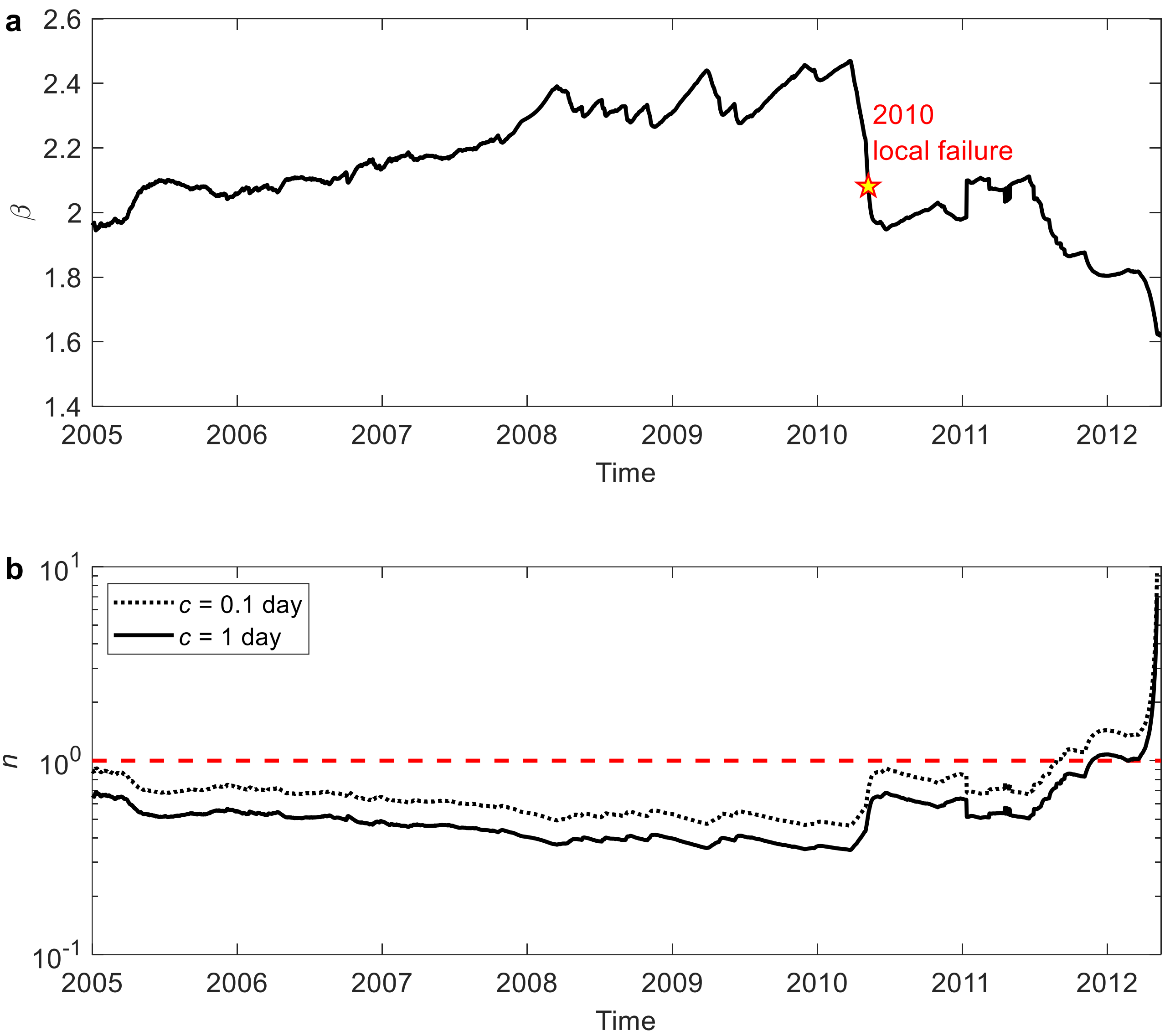

**Supplementary Fig. 11 Temporal variation of the $\beta$-value and $n$-value of the Preonzo landslide. a** Temporal evolution of the power law tail exponent $\beta$ of the velocity probability distribution estimated based on the profile maximum likelihood estimation method; only the $\beta$ evolution after 2005 is shown because sufficient data are needed to constrain the velocity probability distribution; the pentagram marker indicates the 2010 local failure event, prior to which a notable decline of the $\beta$-value can also be observed. **b** Temporal evolution of the branching ratio $n$ (see Supplementary Note 3 for the calibration procedures). The dashed red horizonal line in **b** marks the critical condition of $n = 1$.

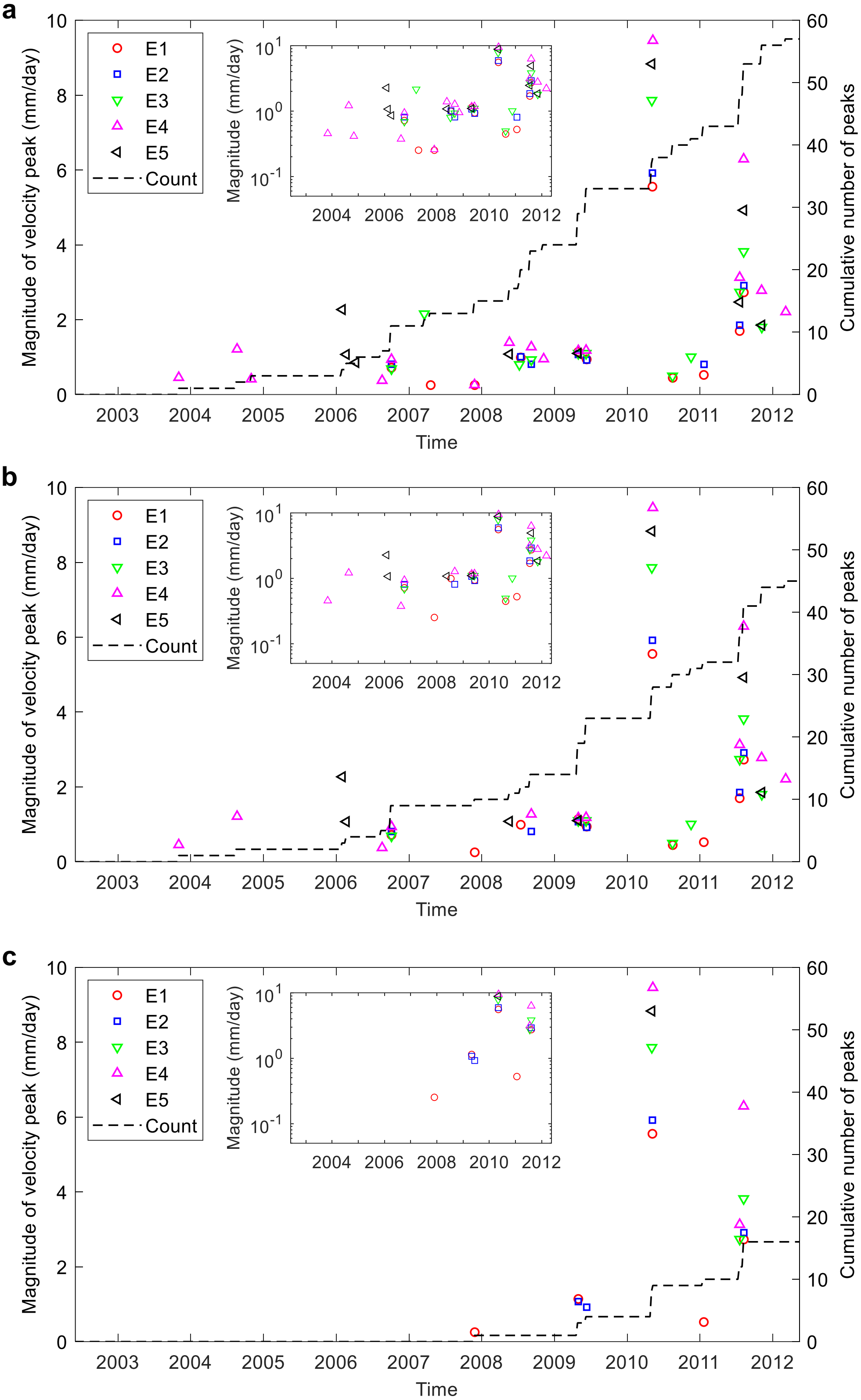

**Supplementary Fig. 12 Temporal evolution of the magnitude and cumulative number of exogenous velocity peaks.** The peaks are selected from those shown in Supplementary Fig. 17 with an additional criterion that the coefficient of determination (in the power law fitting) **a** $R^2 > 0.7$ (see Supplementary Fig. 22a), **b** $R^2 > 0.8$ (see Fig. 6), and **c** $R^2 > 0.9$ (see Supplementary Fig. 22b).

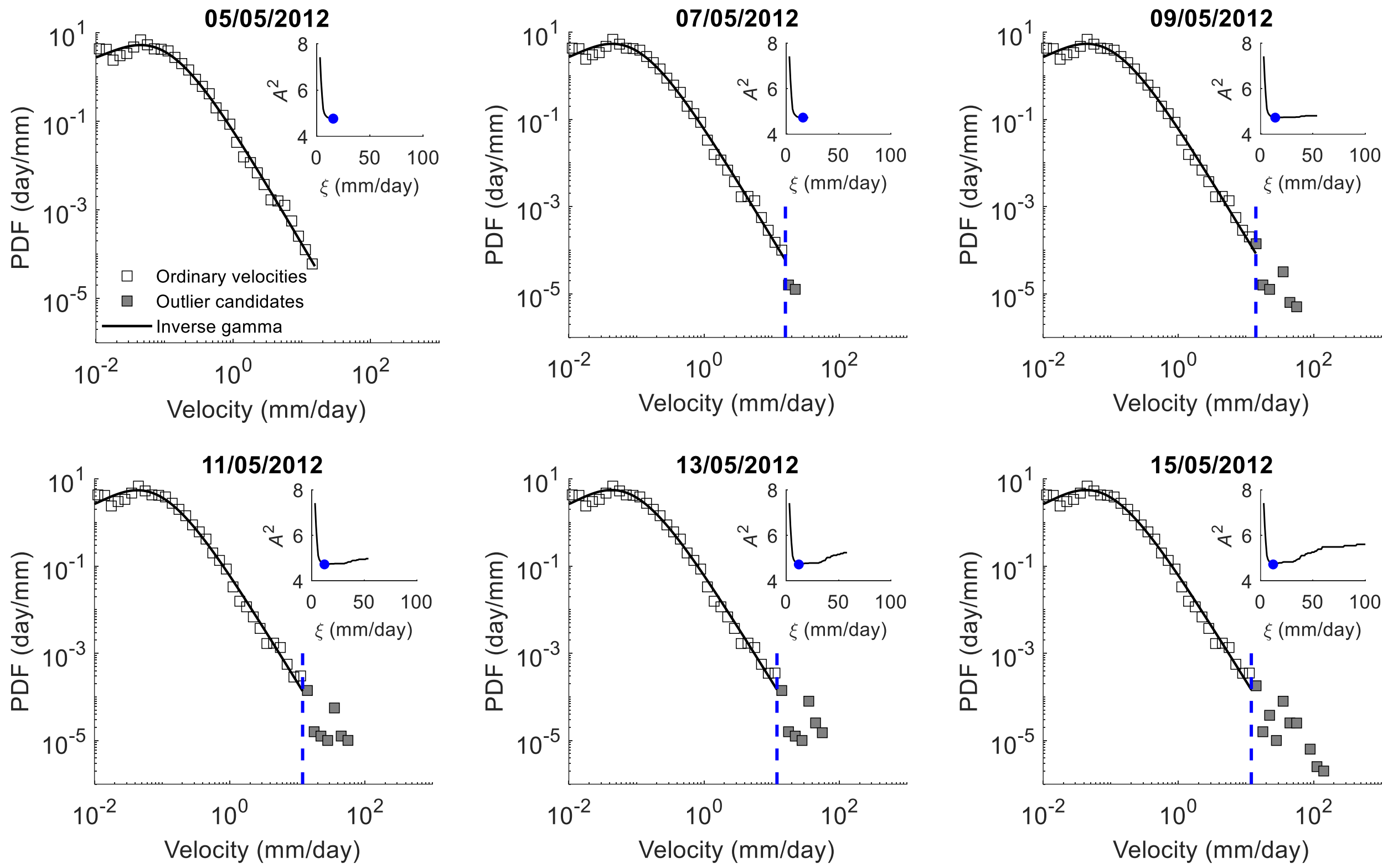

**Supplementary Fig. 13 Probability density functions (PDFs) of daily velocities of the Preonzo landslide on different selected dates close to the 2012 failure event.** Here, the data from all the five extensometers are blended when constructing the PDF. The solid line indicates the inverse gamma distribution fitted by the velocity data excluding outlier candidates (i.e., those larger than the threshold indicated by the blue dashed line if present). Inset: Anderson-Darling test result for determining the threshold velocity $\xi$, beyond which velocities are considered as outlier candidates; the estimate of $\xi$ corresponds to the point (marked by a blue-colored circle) where the Anderson-Darling distance $A^2$ reaches the global minimum. See Supplementary Note 5 for the detailed mathematical formulations.

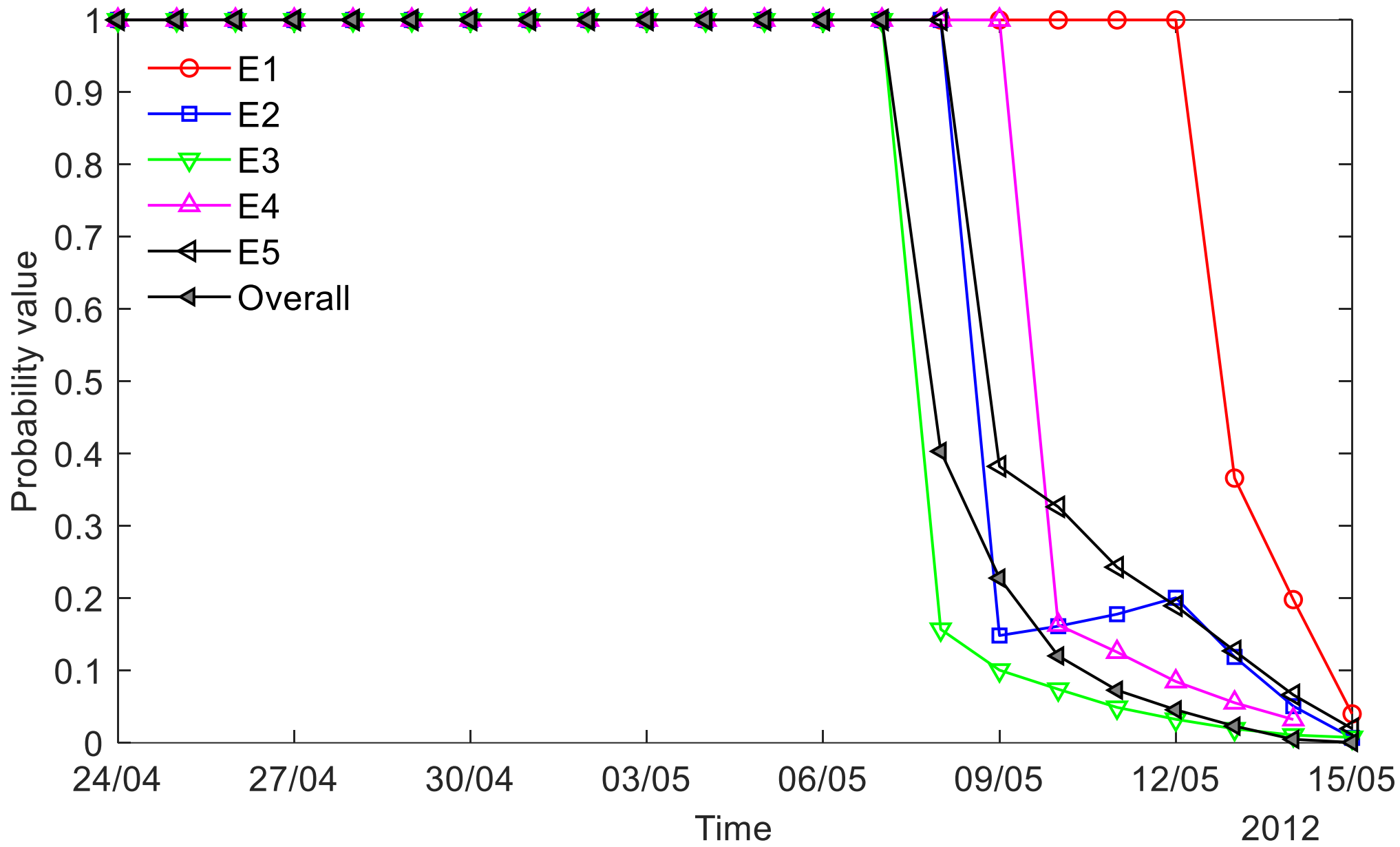

**Supplementary Fig. 14** Temporal evolution of the probability values derived from the dragon-king detection statistical test (see Supplementary Note 5) on the slope velocity data measured by five extensometers. A systematic decline of the probability values after 7 May 2012 signifies the transition of the slope into the dragon-king regime.

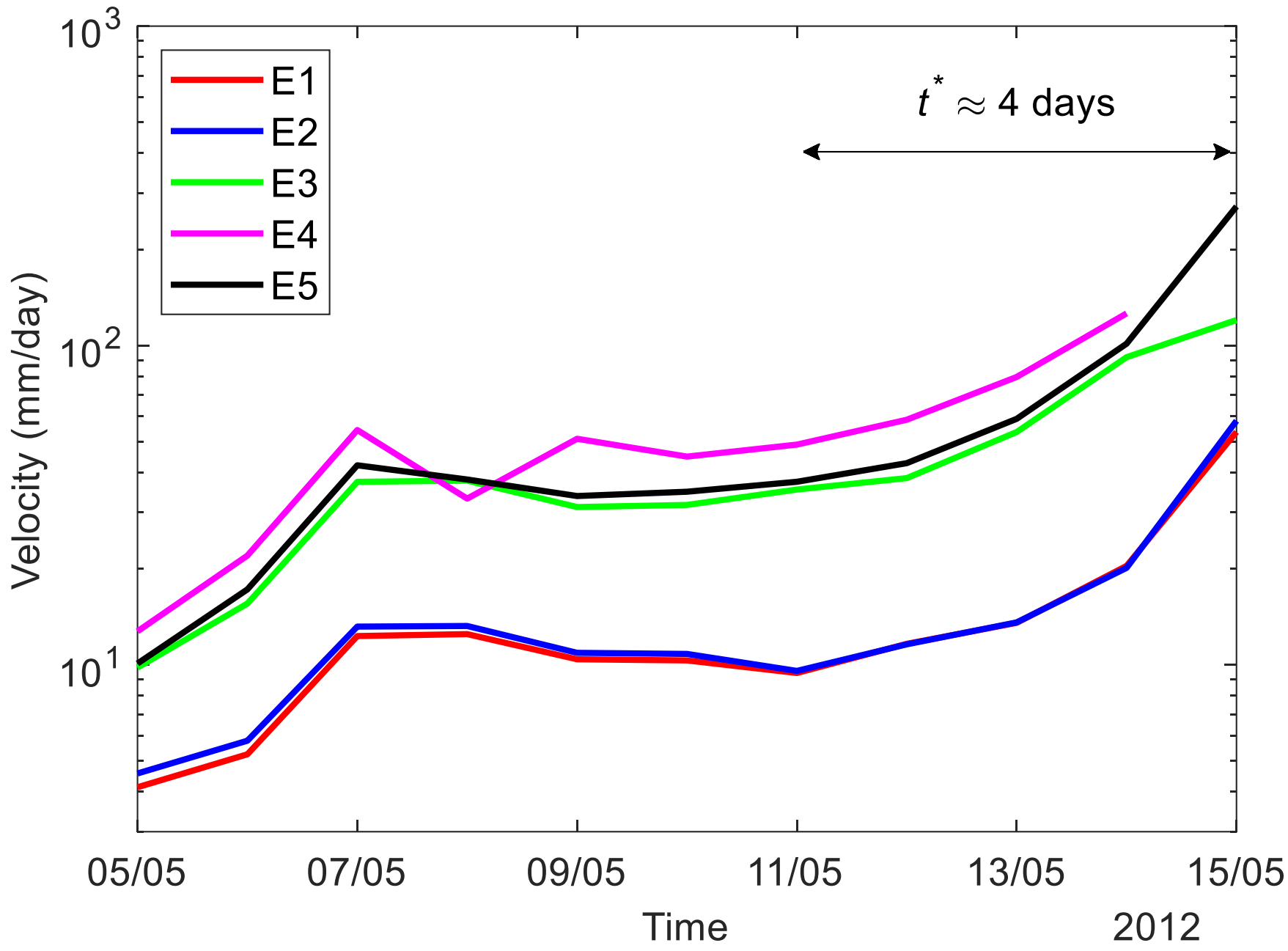

**Supplementary Fig. 15 Precursory deceleration of the Preonzo landslide prior to its catastrophic failure on 15 May 2012.** Here, the characteristic time $t^*$ marks the transition from an early-time deceleration (from 7 to 11 May) to a late-time acceleration (from 11 May to 15 May).

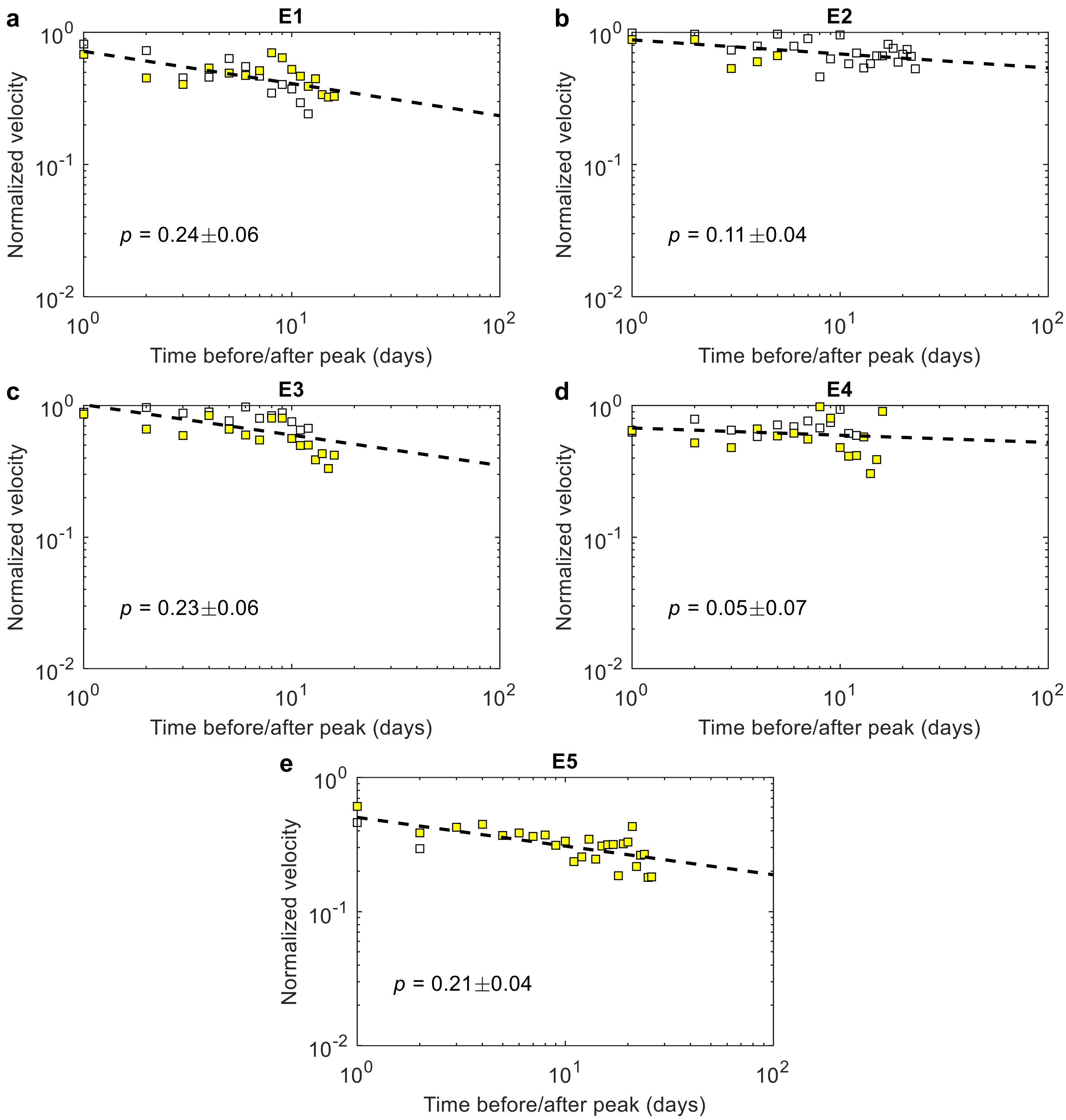

**Supplementary Fig. 16 Pre-peak (open symbols) acceleration and post-peak (colored symbols) relaxation of Type IV endogenous-critical peaks.** Variation of normalized velocity as a function of pre/post-peak time for the five extensometers E1-E5.

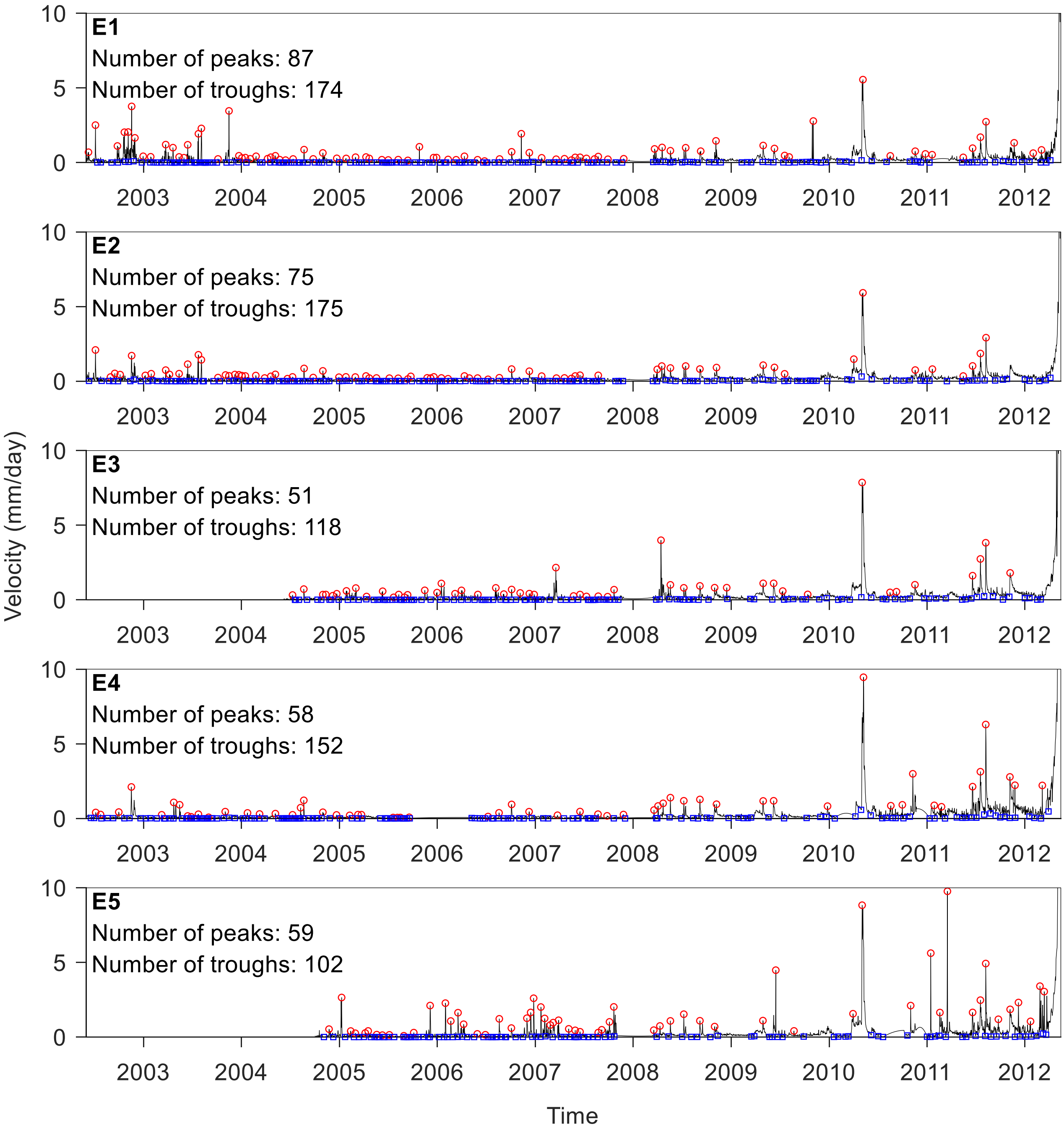

**Supplementary Fig. 17 Time series of daily slope velocities recorded by the five extensometers E1-E5 (from top to bottom) instrumented at the Preonzo landslide, Switzerland.** Peaks and troughs are marked by red circles and blue squares, respectively. Each peak (respectively trough) is qualified as a local maximum (respectively minimum) over a 20-day time window which is at least $k = 2.5$ times larger (respectively smaller) than the average velocity over a 2-month time window.

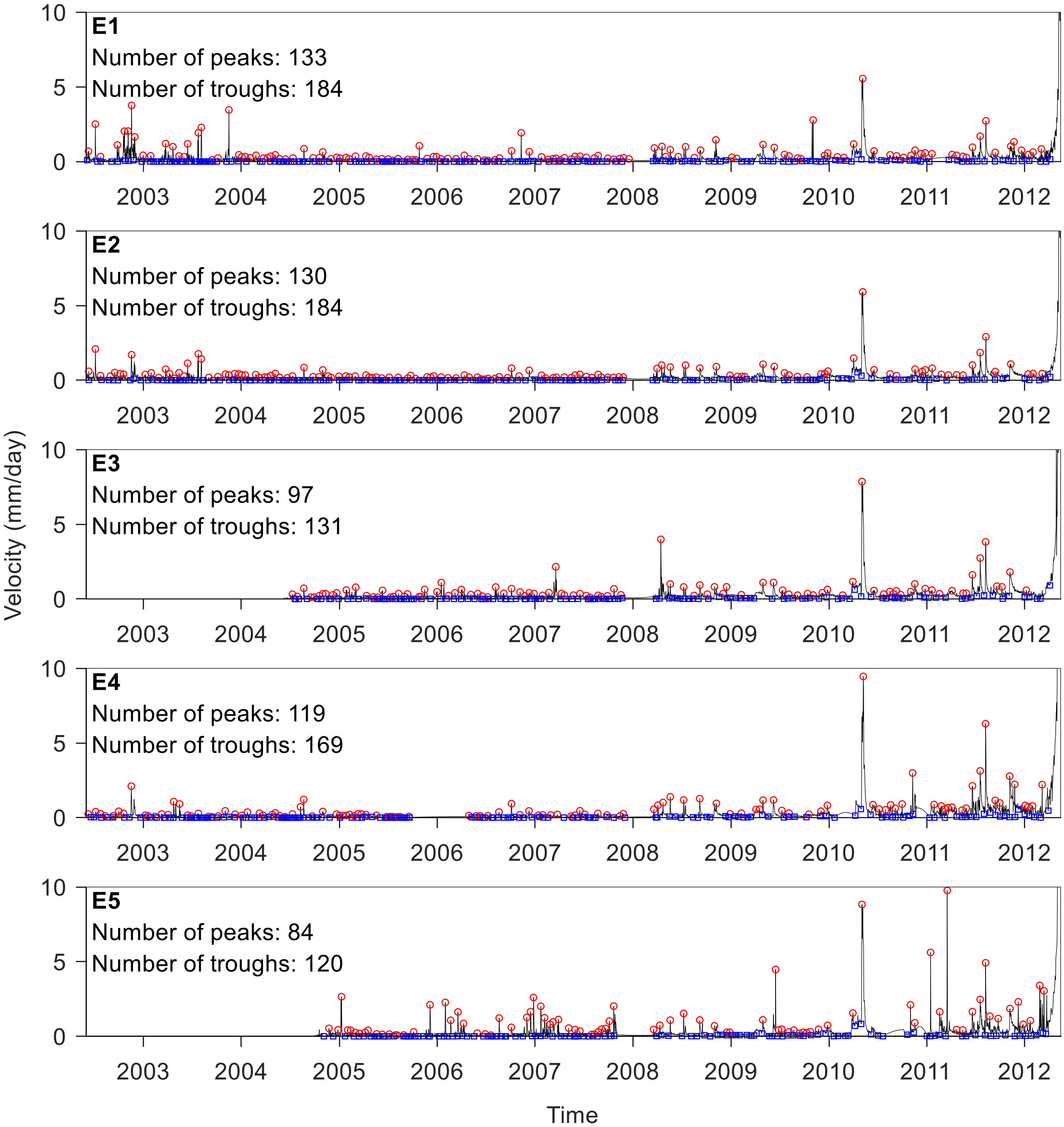

**Supplementary Fig. 18 Time series of daily slope velocities recorded by the five extensometers E1-E5 (from top to bottom) instrumented at the Preonzo landslide, Switzerland.** Peaks and troughs are marked by red circles and blue squares, respectively. Each peak (respectively trough) is qualified as a local maximum (respectively minimum) over a 20-day time window which is at least $k = 1.5$ times larger (respectively smaller) than the average velocity over a 2-month time window.

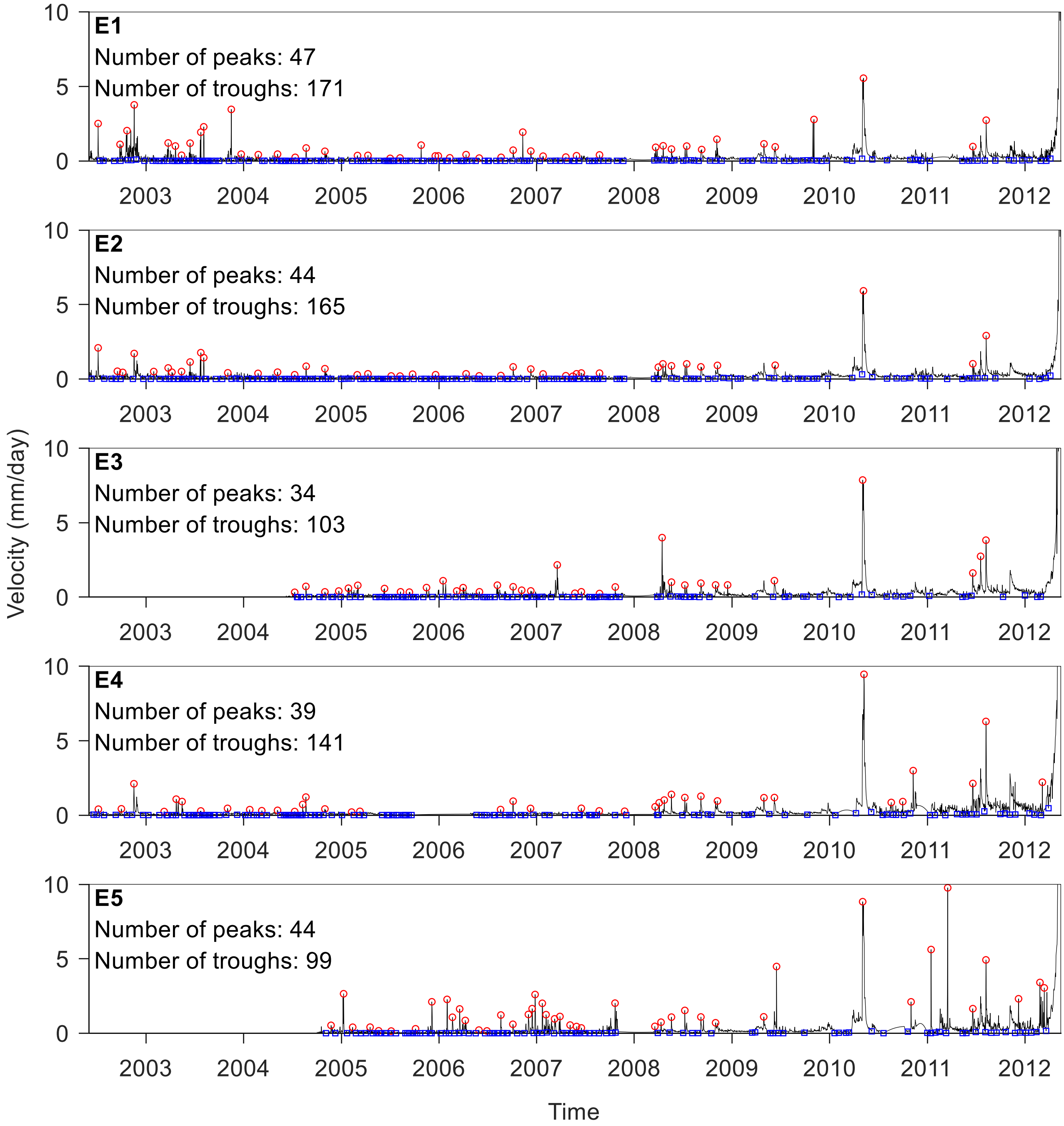

**Supplementary Fig. 19 Time series of daily slope velocities recorded by the five extensometers E1-E5 (from top to bottom) instrumented at the Preonzo landslide, Switzerland.** Peaks and troughs are marked by red circles and blue squares, respectively. Each peak (respectively trough) is qualified as a local maximum (respectively minimum) over a 20-day time window which is at least $k = 3.5$ times larger (respectively smaller) than the average velocity over a 2-month time window.

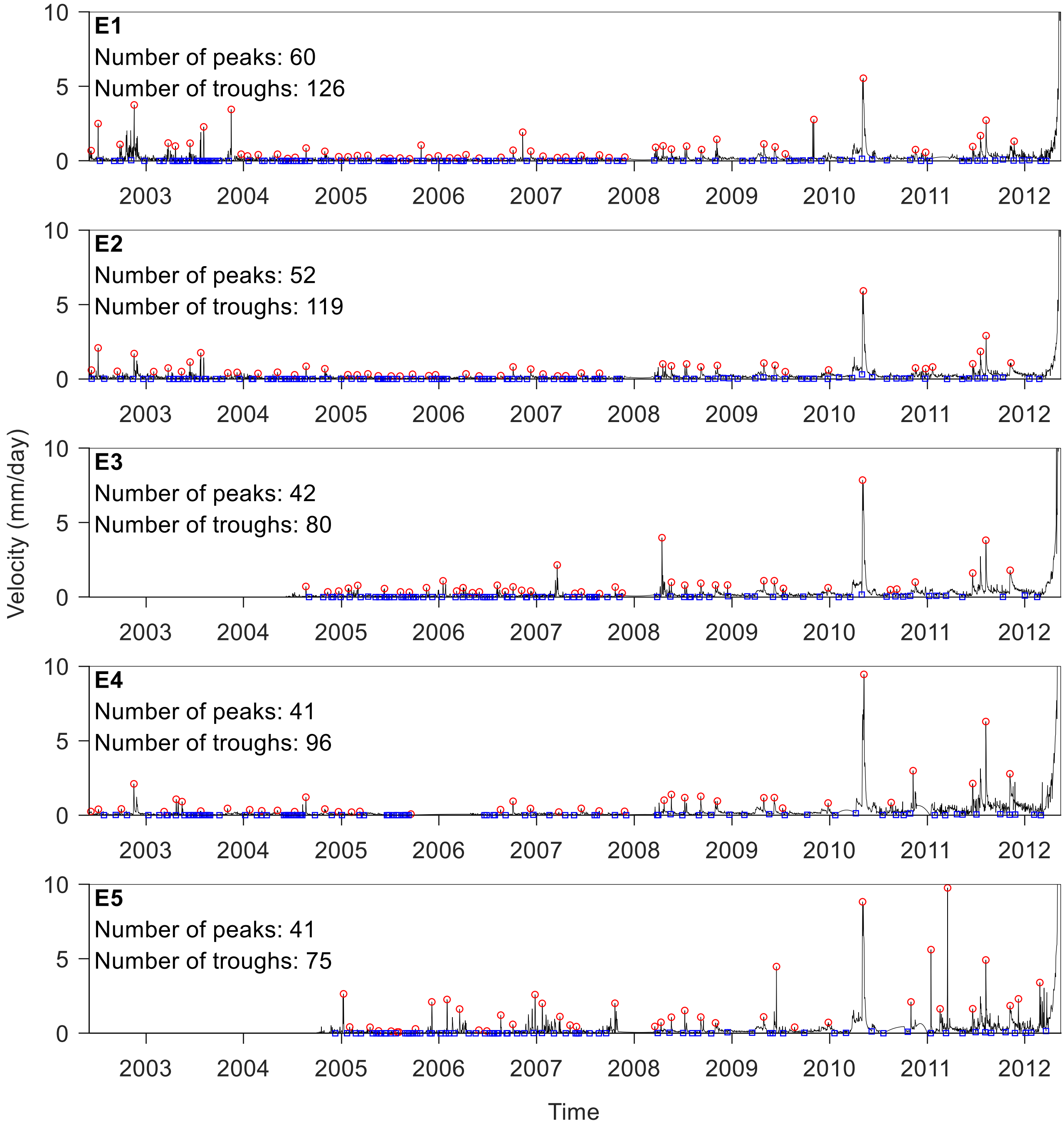

**Supplementary Fig. 20 Time series of daily slope velocities recorded by the five extensometers E1-E5 (from top to bottom) instrumented at the Preonzo landslide, Switzerland.** Peaks and troughs are marked by red circles and blue squares, respectively. Each peak (respectively trough) is qualified as a local maximum (respectively minimum) over a 40-day time window which is at least $k = 2.5$ times larger (respectively smaller) than the average velocity over a 4-month time window.

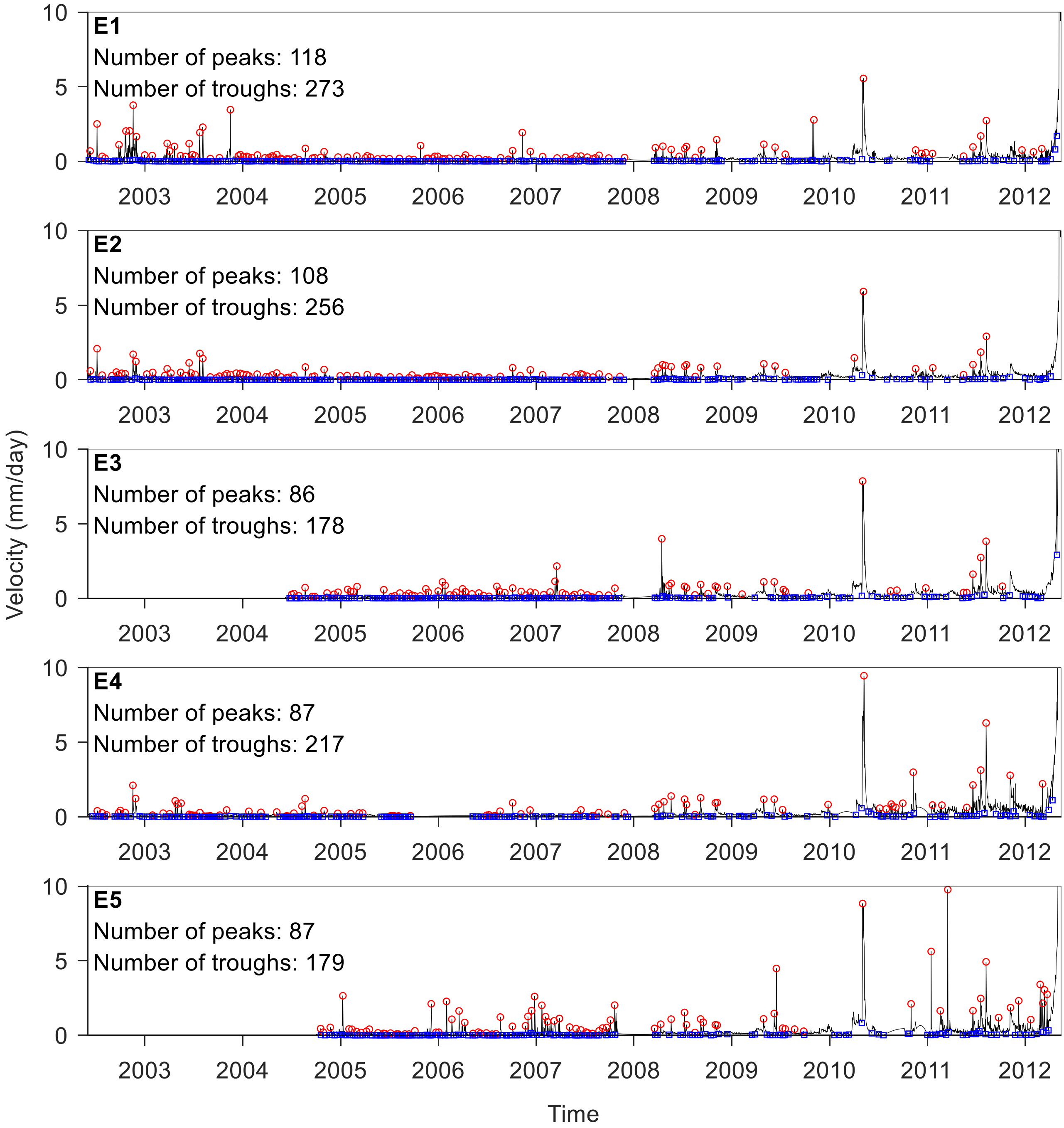

**Supplementary Fig. 21 Time series of daily slope velocities recorded by the five extensometers E1-E5 (from top to bottom) instrumented at the Preonzo landslide, Switzerland.** Peaks and troughs are marked by red circles and blue squares, respectively. Each peak (respectively trough) is qualified as a local maximum (respectively minimum) over a 10-day time window which is at least $k = 2.5$ times larger (respectively smaller) than the average velocity over a 1-month time window.

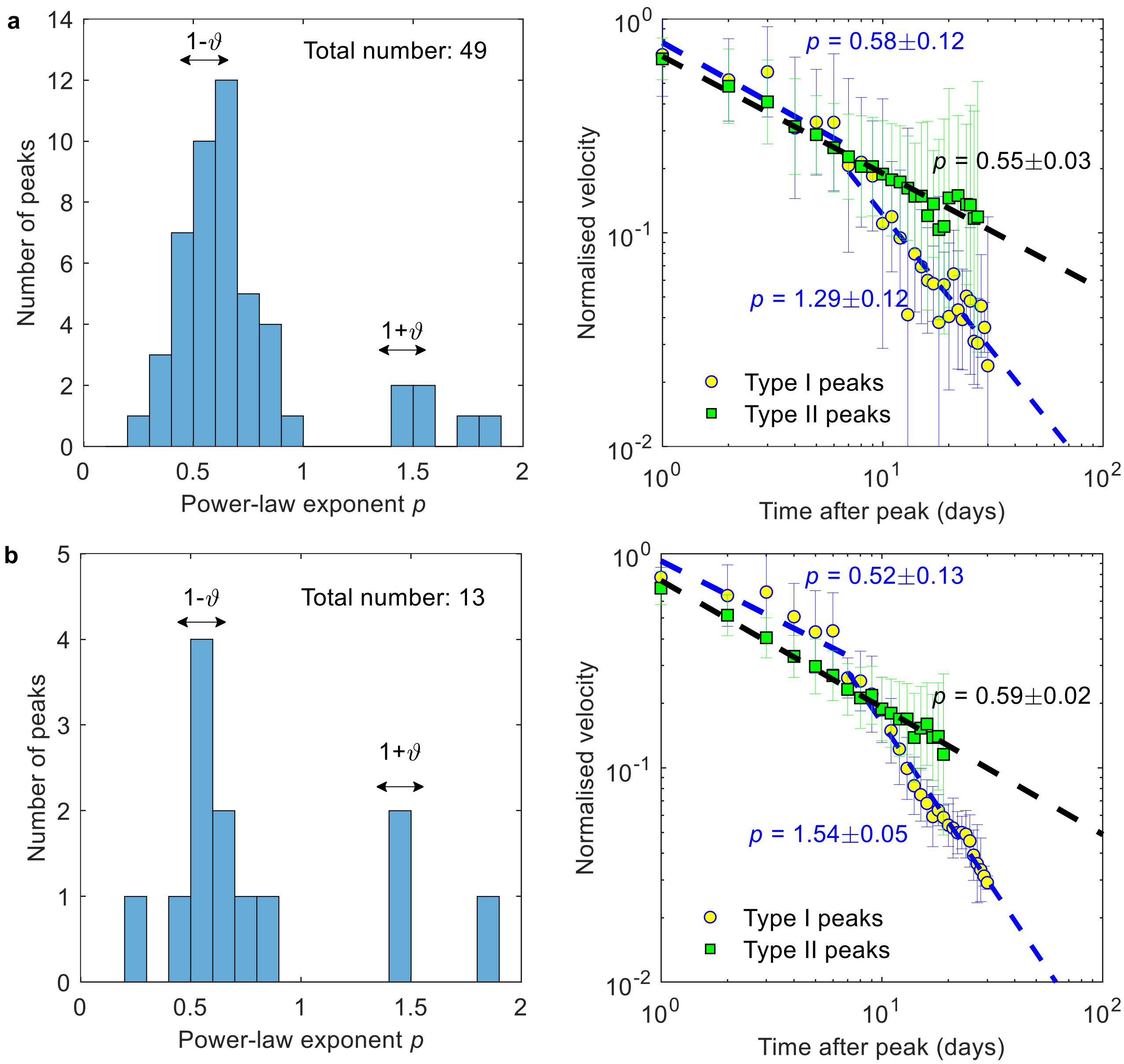

**Supplementary Fig. 22 Post-peak relaxation properties associated with detected peaks in the velocity time series.** Left: histogram of the power law exponents $p$ for post-peak velocity relaxation. Right: ensemble averaged relaxation of Type I (exogenous-subcritical) and Type II (exogenous-critical) peaks. Here, a peak is qualified as a local maximum over a 20-day time window which is at least $k = 2.5$ times larger than the average velocity over a 2-month time window, while the coefficient of determination for the fitting should meet **a** $R^2 > 0.7$ (upper panel) or **b** $R^2 > 0.9$ (lower panel).

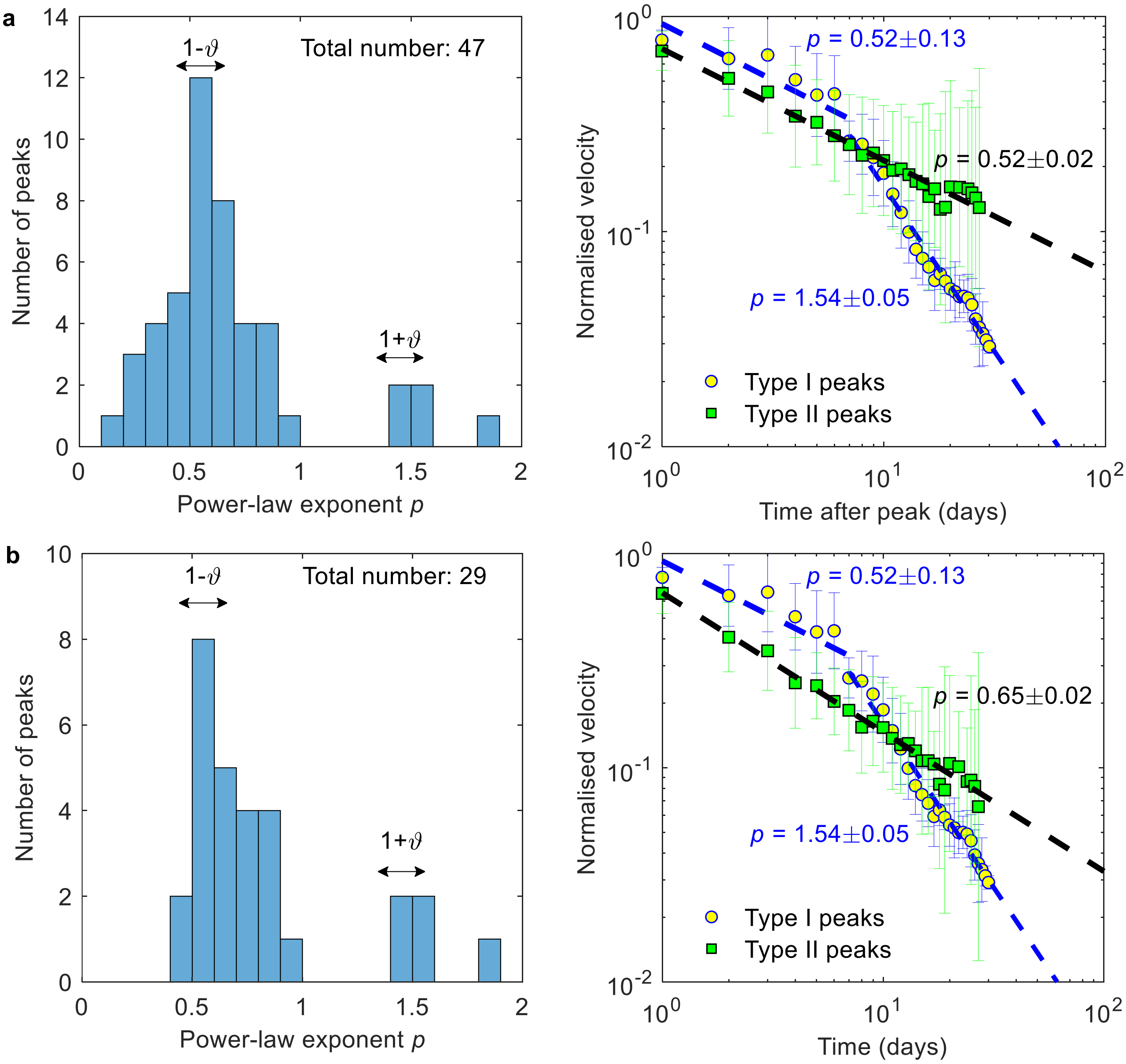

**Supplementary Fig. 23 Post-peak relaxation properties associated with detected peaks in the velocity time series.** Left: histogram of the power law exponents $p$ for post-peak velocity relaxation. Right: ensemble averaged relaxation of Type I (exogenous-subcritical) and Type II (exogenous-critical) peaks. Here, a peak is qualified as a local maximum over a 20-day time window which is at least **a** $k = 1.5$ (upper panel) or **b** $k = 3.5$ (lower panel) times larger than the average velocity over a 2-month time window, while the coefficient of determination for the fitting should meet $R^2 > 0.8$.

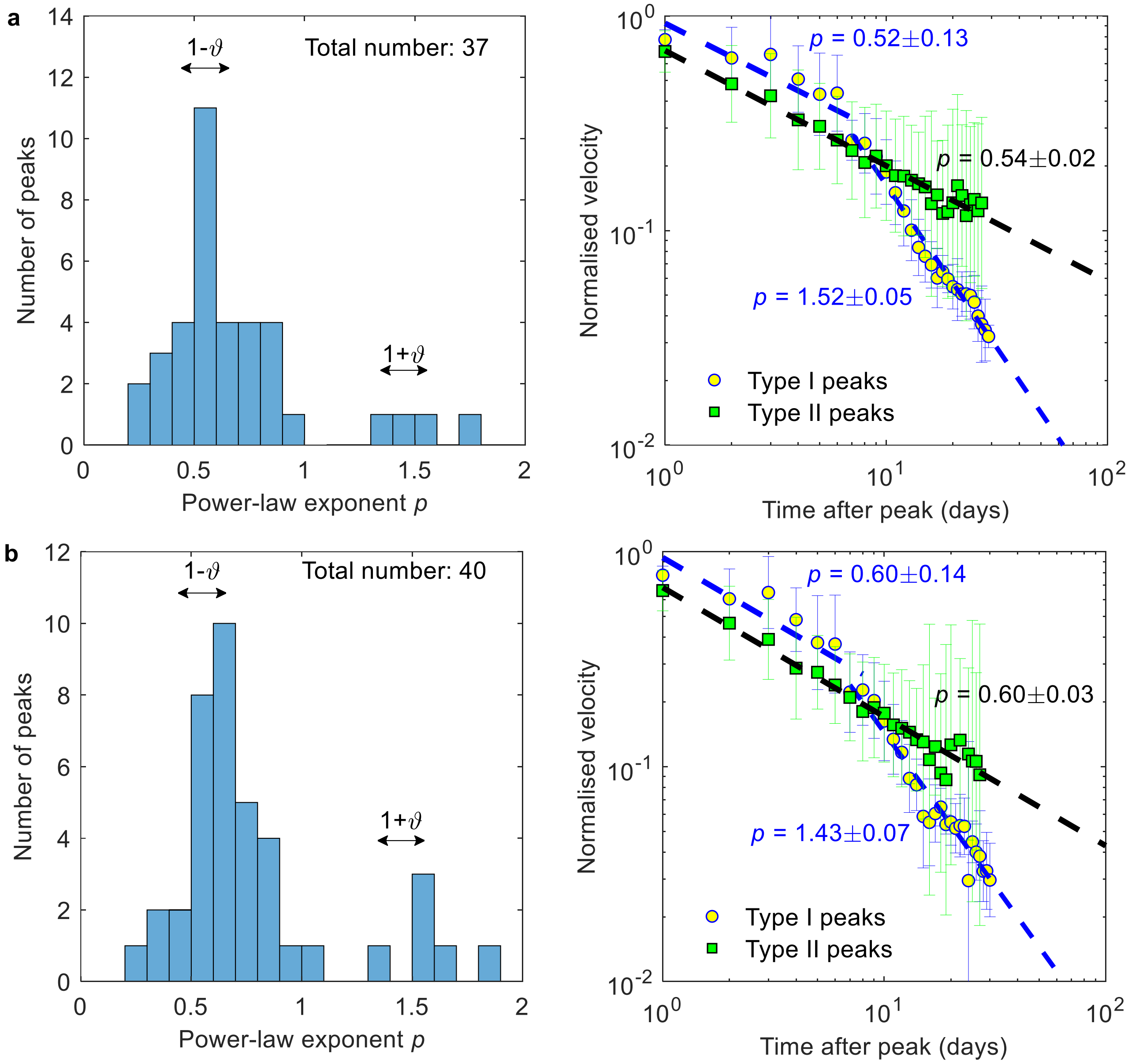

**Supplementary Fig. 24 Post-peak relaxation properties associated with detected peaks in the velocity time series.** Left: histogram of the power law exponents $p$ for post-peak velocity relaxation. Right: ensemble averaged relaxation of Type I (exogenous-subcritical) and Type II (exogenous-critical) peaks. Here, a peak is qualified as a local maximum over a **a** 40-day window which is at least $k = 2.5$ times larger than the average velocity over a 4-month (upper panel) or **b** 10-day time window which is at least $k = 2.5$ times larger than the average velocity over a 1-month time window (lower panel), while the coefficient of determination for the fitting should meet $R^2 > 0.8$.

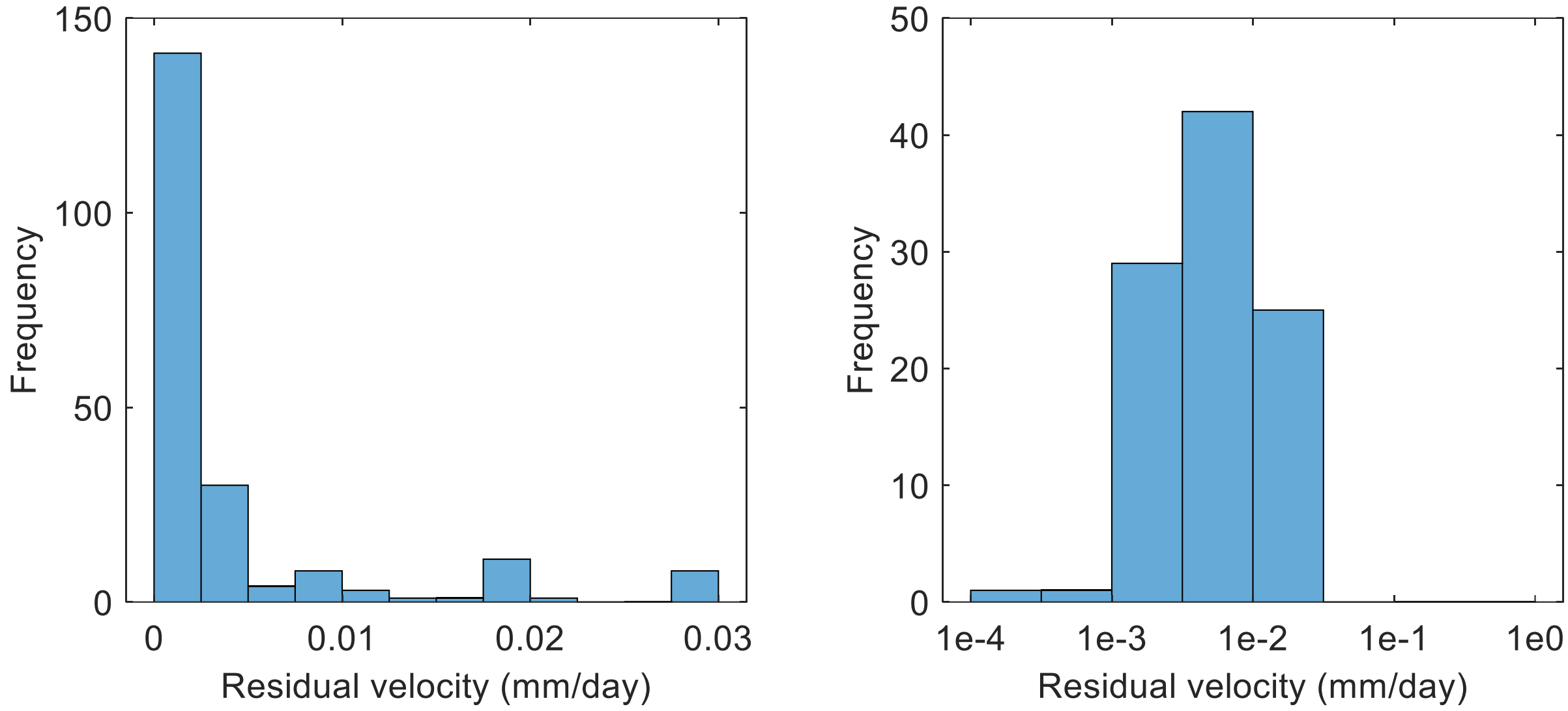

**Supplementary Fig. 25 Histogram of slope residual velocities.** Data are plotted in a linear scale (left) and a logarithmic scale (right).

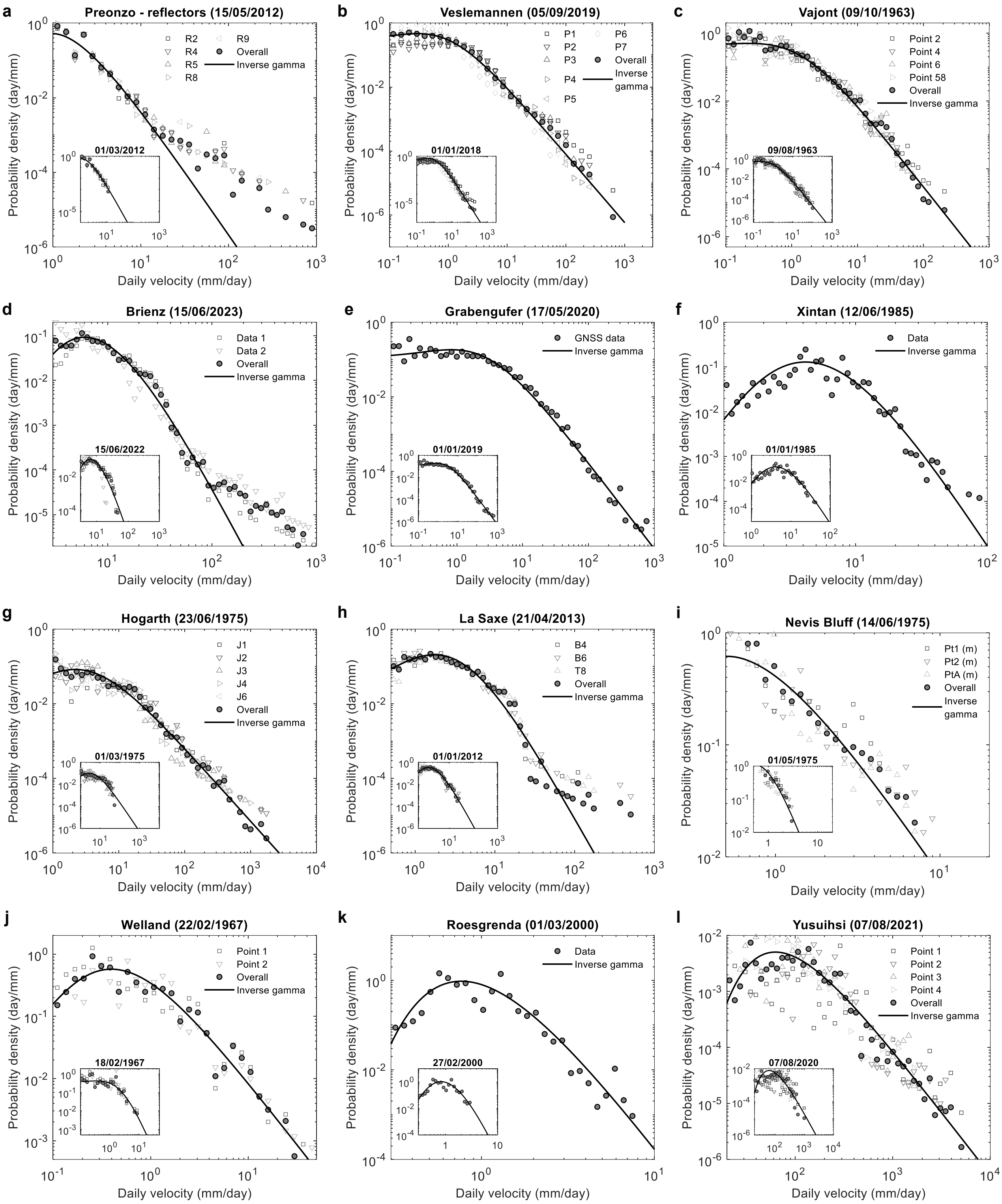

**Supplementary Fig. 26 Probability density function of slope velocities fitted to the inverse gamma function for various landslides.**

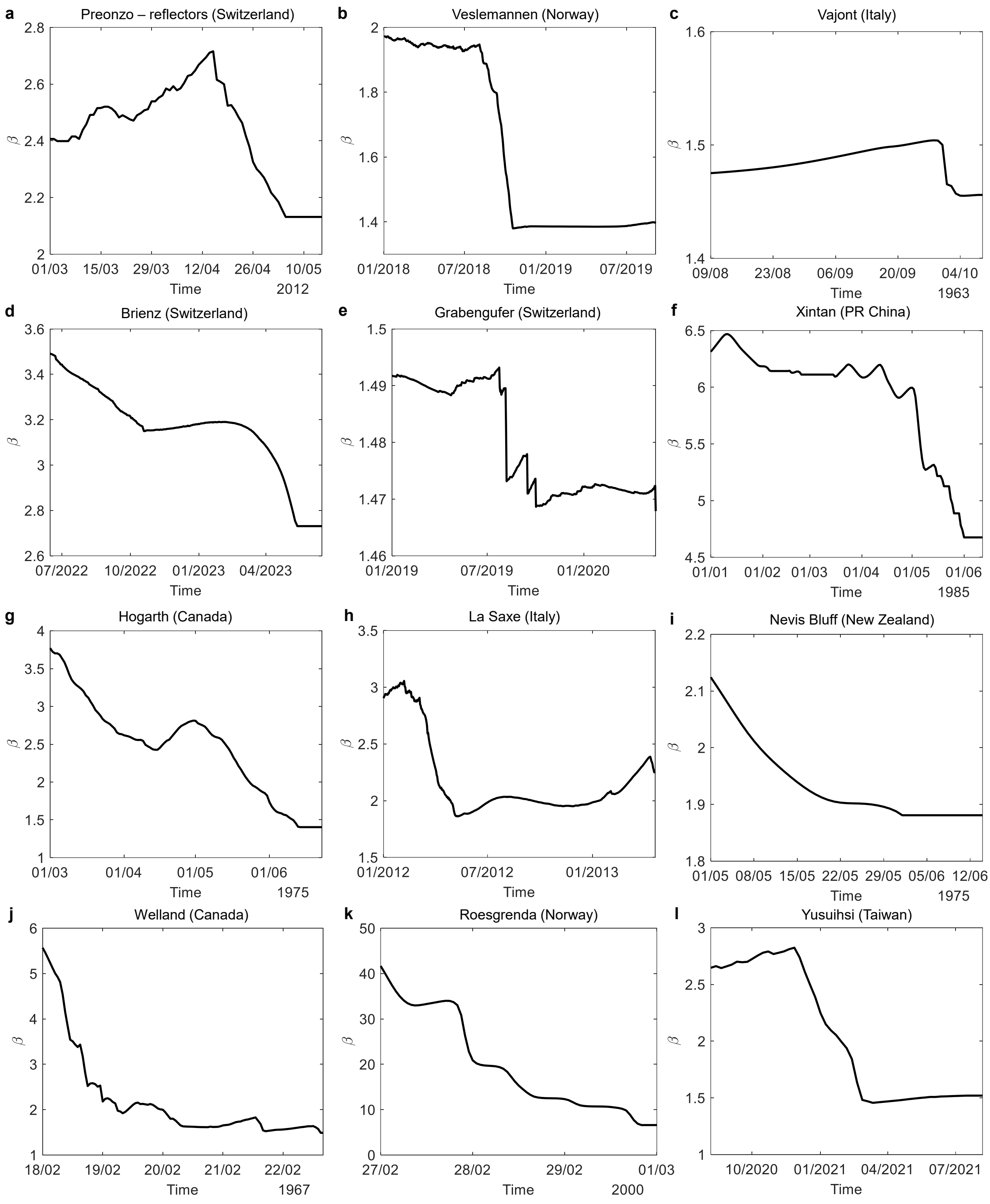

**Supplementary Fig. 27 Progressive decrease with time of exponent *β* of the slope velocity probability distribution preceding catastrophic failure in various landslides.**

**Supplementary Table 1 Landslide information (25 landslides in total).**

| Site | Location | Type | Material | Failure time | Volume ($m^3$) | Monitoring method |
|---|---|---|---|---|---|---|
| Brienz[30] | Switzerland | Rockslide | Flysch, schists, dolomite | 2023-06-15 | $1.2\times10^6$ | Total station with reflectors |
| Grabengufer[31] | Switzerland | Rockfall | Rock & ice | 2020-05-17 | $5\times10^2$ | GNSS & inclinometer |
| Hogarth[32] | Canada | Topple | Diorite | 1975-06-23 | $2\times10^5$ | Extensometers |
| Hollin Hill[8] | UK | Rockslide | Sandstone, limestone, mudstone & siltstone | N/A | $\sim5\times10^3$ | Shape acceleration arrays |
| Ingelsberg[12] | Austria | Rockfall | Schists | 2013-04-29 | 20-40 | Ground-based InSAR |
| La Saxe[3] | Italy | Rockslide | Meta-sedimentary sequences | 2013-04-21 | $5\times10^2$-$1\times10^3$ | Total station with reflectors |
| La Clapière[16] | France | Rockslide | Metamorphic rocks | N/A | $5\times10^7$ | Distance meters |
| Maca[5] | Peru | Soilslide | Fine-grained lacustrine sediments | N/A | $6\times10^7$ | Continuous GPS stations |
| Moosfluh[9] | Switzerland | Rockslide | Metamorphic rocks | N/A | $7.5\times10^7$ | Total station with reflectors |
| Nevis Bluff[33] | New Zealand | Flexural topple / rockslide | Schist | 1975-06-14 | $3.2\times10^4$ | Survey markers |
| Oak Ridge[11] | USA | Earthflow | Fragments of sandstone, chert, greenstone, and blueschist | N/A | $\sim2\times10^6$ | Extensometers |
| Pomarico[4] | Italy | Soilslide / earthflow | Cays & sands | N/A | $4\times10^6$ | Ground-based InSAR |
| Preonzo[34] | Switzerland | Rockslide | Gneiss | 2012-05-15 | $2.1\times10^5$ | Extensometers & total station with reflectors |
| Roesgrenda[35] | Norway | Soilslide | Quick clay | 2000-03-02 | $2\times10^3$ | Extensometers |
| Séchilienne[13] | France | Rockslide | Micaschists | N/A | $\sim4\times10^6$ | Extensometers |
| Shuping[10] | PR China | Soilslide | Colluvial rubble soils & residual clayey soils | N/A | $1.6\times10^7$ | Continuous GPS stations |
| Tapgaon[6] | Nepal | Rockslide | Regolith & weathered rocks | N/A | $\sim1\times10^8$ | Pléiades optical satellite images and SAR Sentinel-1 images |

**Supplementary Table 1 (continued) Landslide information (25 landslides in total).**

| Site | Location | Type | Material | Failure time | Volume ($m^3$) | Monitoring method |
|---|---|---|---|---|---|---|
| Twain Harte[2] | USA | Exfoliation dome (related to rockfalls) | Granite | N/A | $2.3\times10^3$ | Extensometers |
| Vajont[36] | Italy | Rockslide | Limestone | 1963-10-09 | $2.7\times10^8$ | Geodetic bench marks |
| Vallcebre[37] | Spain | Rockslide | Shale, gypsum & claystone | N/A | $\sim2\times10^7$ | Borehole wire extensometers |
| Veslemannen[1] | Norway | Rockslide | Gneiss | 2019-09-05 | $5.4\times10^4$ | Ground-based InSAR |
| Welland[38] | Canada | Soilslide | Clay | 1967-02-22 | $5\times10^2$ | Extensometers |
| Xintan[39] | PR China | Rockslide | Sediments | 1985-06-12 | $3\times10^7$ | Geodetic bench marks |
| Yaoshan[7] | PR China | Soilslide | Colluvium and residual clay | N/A | $5\times10^5$ | Inclinometers |
| Yusuihsi[40] | Taiwan | Rockslide (evolved into debris flow after failure) | Slate & phyllite | 2021-08-07 | $8.5\times10^6$ | Optical satellite imagery |

## Supplementary References